\documentclass[prd,aps,a4paper,superscriptaddress,twocolumn,nofootinbib]{revtex4}
\usepackage{graphicx}
\usepackage{color}
\usepackage{dcolumn}
\usepackage{bm}
\usepackage{slashed}
\usepackage{amsmath}
\usepackage{latexsym}
\usepackage{amssymb}
\usepackage{amsfonts}
\usepackage{url}
\allowdisplaybreaks
\begin{document}
\title{Gravitational wave signal recognition of O1 data by deep learning}
\author{He Wang}
\affiliation{Department of Physics, Beijing Normal University, Beijing 100875, China}
\author{Zhoujian Cao\footnote{corresponding author}} \email[Zhoujian Cao: ]{zjcao@amt.ac.cn}
\affiliation{Department of Astronomy, Beijing Normal University, Beijing 100875, China}
\author{Xiaolin Liu}
\affiliation{Department of Astronomy, Beijing Normal University, Beijing 100875, China}
\author{Shichao Wu}
\affiliation{Department of Astronomy, Beijing Normal University, Beijing 100875, China}
\author{Jian-Yang Zhu}
\affiliation{Department of Physics, Beijing Normal University, Beijing 100875, China}

\begin{abstract}
Deep learning method develops very fast as a tool for data analysis these years. Such a technique is quite promising to treat gravitational wave detection data. There are many works already in the literature which used deep learning technique to process simulated gravitational wave data. In this paper we apply deep learning to LIGO O1 data. In order to improve the weak signal recognition we adjust the convolutional neural network (CNN) a little bit. Our adjusted convolutional neural network admits comparable accuracy and efficiency of signal recognition as other deep learning works published in the literature. Based on our adjusted CNN, we can clearly recognize the eleven confirmed gravitational wave events included in O1 and O2. And more we find about 2000 gravitational wave triggers in O1 data.
\end{abstract}

\maketitle

%\tableofcontents

\section{Introduction}
Gravitational waves (GW) are an important prediction of Einstein's general theory of relativity, published a century ago. Gravitational wave observations of coalescing compact binaries are unique and unprecedented probes of strong-field and dynamical aspects of general relativity. More importantly, it ushers the brand new cosmic window to our universe---gravitational wave astronomy \cite{PhysRevLett.116.061102,PhysRevLett.116.241103,PhysRevX.6.041015,PhysRevLett.118.221101,PhysRevLett.119.141101,PhysRevLett.119.161101,2041-8205-851-2-L35,LIGOScientific:2018mvr}.

The capability of searching for GW signals relies on both the sensitivity of GW detectors and the theoretical waveform templates modeled for gravitational wave sources which will be used by matched-filtering data analysis technique. Currently the matched-filtering method is the standard and optimal signal processing techniques used by the gravitational wave community to find GW signals from compact BBH mergers in noisy detector data. Although the weak signal extraction and source information inversion of GW based on matched-filtering techniques are very successful, it has a great weakness as well as a potential hazard: data analysis through matched-filtering is a process of both a huge computation cost and slow computational speed. This is the major motivation for many authors to propose deep learning technique for GW data analysis \cite{PhysRevD.97.044039,GEORGE201864,2017arXiv171200356L,fanli18}. Another possible problem with standard matched-filtering techniques is that the completeness and accuracy of GW waveform template is the prerequisite to guarantee its work. This implies a risk that we may lose the GW signals beyond the theoretical expectation. But the GW signal beyond the expectation of the theory will greatly facilitate the development of astronomy, and also provide important insight to the problems of fundamental physics such as quantum gravity and physics under extreme conditions.

The matched-filtering provides an optimal solution for identification of gravitational waves under Gaussian noise, but in practice, the data from GW detectors contains many non-Gaussian noise transients, also known as `glitches'. A comprehensive classification and characterization of these noise features may provide valuable clues for identifying the source of noise transients, and possibly lead to their elimination. The machine learning method is becoming more and more important in various disciplines such as particle experiment \cite{Cohen2018,PhysRevD.97.056009}, gamma ray detection \cite{1475-7516-2018-05-058,abraham2019machine}, supernovae classification \cite{charnock2017deep,Moss:2018tug}, weak lensing data analysis \cite{gupta2018non,Shirasaki:2018thk,Ribli:2019wtw,Fluri:2019qtp}, source modeling \cite{Chua:2018woh,Rebei:2018lzh,Chua:2019wwt,Setyawati:2019xzw} and others. There have also been many attempts to use machine learning algorithms in gravitational wave data analysis to show promise for noise classification, categorization \cite{RAMPONE:2013oga,Mukherjee:2010zza,Powell:2015ona,Powell:2016rkl,PhysRevD.97.101501,PhysRevD.95.104059,0264-9381-35-9-095016,2017CQGra..34f4003Z,George:2017fbn}, and cancellation \cite{Wei:2019zlc,Shen:2019ohi}.
And recently, an innovative project called ``Gravity Spy"\footnote{www.gravityspy.org} \cite{0264-9381-34-6-064003,BAHAADINI2018172} combines the power of machine learning with the help of volunteers to label datasets of glitches and create a superior classifier of glitches in LIGO data. Machine learning techniques have been widely used in GW data processing, especially in the identification of signals and the classification of noises.

In recent years, deep learning, a new area of machine learning research, has been in the spotlight \cite{Allen:2019dkq}. In the past few years, some active researchers have demonstrated empirical successes of deep neural networks in the applications to data analysis of gravitational waves \cite{Gabbard:2019rde,Chatterjee:2019gqr,Miller:2019jtp,Krastev:2019koe,Shen:2019vep}. These active researchers include George and his coworkers \cite{PhysRevD.97.044039,GEORGE201864}, Gabbard and his coworkers \cite{PhysRevLett.120.141103}, and others \cite{2017arXiv171200356L,fanli18,Wang:2019ybv}. These published works used the convolutional neural network (CNN) from different perspectives to identify GW signals with low signal-to-noise ratio (SNR). Their works tell us that CNN architecture plays an important role for CBC GWs (also continuous GWs \cite{PhysRevD.100.044009,2019arXiv190706917M} and CCSNe GWs \cite{PhysRevD.98.122002}) recognition in the simulated or real noisy background from LIGO. Whereas, Gebhard et al. \cite{Gebhard:2019ldz} have given an enlightening discussion on the general limitations of CNNs and proposed an alternative CNN-based architecture with proper performance metrics. They also claim that their trained network can cover all the GW events in both the first and second observation run (O1/O2) of LIGO, only except for GW170817, the first observation of GW from a binary neutron star inspiral.

In this paper we aim to use deep learning technique to find out all of the known GW events in the O1 and O2 data. In addition we would like to mark out other possible GW events candidates. We have tried CNN like the ones used in \cite{PhysRevD.97.044039,GEORGE201864,2017arXiv171200356L,fanli18}. Although these neural networks can find GW150914, none of them can find out other events in O1. So we adjust the usual CNN a little bit. Based on our adjusted CNN networks, we can find out all 11 GW events reported by LSC \cite{LIGOScientific:2018mvr}. Besides these confirmed events, we have also found out 2242 triggers in O1.

The plan of this paper is as follows: In Sec.~\ref{sec:MFCNN} we describe the adjustment of the CNN neural network. Based on our adjusted CNN architecture, training data samples, test data samples, training strategy and search methodology on the real LIGO recording are described in Sec.~\ref{sec:Method_Strategy}. After that we apply our trained network to the O1 data in Sec.~\ref{sec:result}. At last the Sec.~\ref{sec:Summary} is devoted to a summary.

\section{\label{sec:MFCNN}Adjusted convolutional neural networks}
The inner product operation in frequency domain
\begin{equation}
\langle d|h \rangle = 4\int^\infty_0\frac{\tilde{d}(f)\tilde{h}^*(f)}{S_n(f)}e^{2\pi ift_c+\phi_c}df
\label{eq:corr}
\end{equation}
can also be expressed as a convolution in time domain. We divide the first layer of our adjusted CNN network into $N_t$ groups. The coefficients $h_{ti}, i=1,...,N_t$ of each group are fixed and correspond to a whitened theoretical waveform. After the convolution operation between the input data and each group neural (we leave the detail operation to the App.~\ref{app}), we output the maximal value of each convolution. Then we collect these maximal values as the output of the first layer. The rest layers of our adjusted network is the same as usual CNN network \cite{PhysRevD.97.044039,GEORGE201864}.

The coefficients $h_{ti}$ are analogous to the template waveforms in matched filtering data analysis (Here we use index $t$ to remind template). Intuitively yes, we even chose some template waveforms used by LSC to set these $h_{ti}$. Our basic idea is using these specific template waveforms to sense the GW signal deeply buried in the noise. But different to usual matched filtering technique, we need only tens of templates here instead of millions of templates in usual matched filtering technique. In the current work, we use $N_t=35$ which works quite good to find signal in the O1 data.

Logically our adjusted CNN works in the following way. The coefficients $h_{ti}$ we chosen span a subspace of the function space of the GW signal. After the first layer, the essential matched filtering operation dig out the signal buried in the noise and project it into the subspace. Definitely such projection will admit some feature structure which may not be recognized by human but can be recognized by the following CNN layers. Such feature can be used to distinguish GW signal from pure noise.

Regarding to the 35 templates used in the first layer of our adjusted CNN network, we chose spinless identical binary black holes with total masses $M=5+2i$M${}_\odot,i=0,...,34$.

Before we send the data into our CNN, a tukey window\footnote{\url{en.wikipedia.org/wiki/Window_function#Tukey_window}} with $\alpha=1/8$ is applied to remove edge effects at the beginning and end of the data stretch.
\section{\label{sec:Method_Strategy}Training and test of the neural network}

\subsection{Training data set and test data sets}
\begin{figure}
\includegraphics[width=\columnwidth]{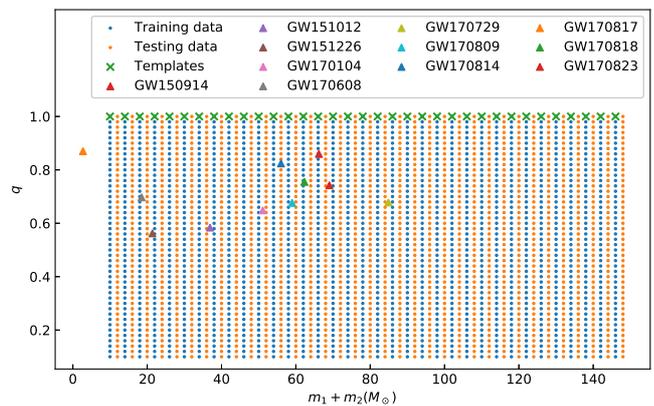}
\caption{\label{fig1}
The total mass and mass ratio of training/test data and templates. The eleven GW events for both O1 and O2 are also shown.}
\end{figure}
Each data sample include noise part and possibly signal part. We obtain the O1 data from the Gravitational Wave Open Science Center (GWOSC) \cite{Vallisneri_2015}. The background noises for training/testing are sampled from O1 data without GW150914, GW151012 and GW151226 events. Together with simulated GW signal we construct 3220 samples for training data and test data respectively. Among these 3220 samples, half of them are pure noise and the rest half include signals.

We use our SEOBNRE model \cite{PhysRevD.96.044028} to generate simulated gravitational waveforms. In the current work we only consider circular, spinless binary black holes. Corresponding to LIGO detectors we adopt the configuration from LALSimulation\footnote{\url{lscsoft.docs.ligo.org/lalsuite/lalsimulation/group__lalsim__detector__strain.html}}  that set all the binary sources at right-ascension 1h 23m 45s, declination 45 degrees, and polarization angle 30 degrees and consider the total mass of the two black holes fall in the range $10\sim150M_\odot$ in steps of 2 and the mass ratio $q=m_1/m_2$ from 1 to 10 in steps of 0.1. Regarding the orbital plane direction we set $\iota=0$. For the source luminosity distance $D$, we determine it through signal-to-noise ratio assumption. Thus, we create a training and testing data set both with 1610 waveforms. Each data sample consists two time-series of 5 seconds long through a sampling rate of 4096 Hz. The two time-series corresponds to detector H1 and L1 respectively. For the samples with GW signal we set the peak location of the signal at the center of the time-series. The mass distribution of templates, training/test waveforms and also the eleven GW events in O1/O2 are plotted in Fig.~\ref{fig1}.

\subsection{Training strategy}

The coefficients of our adjusted CNN network except the first layer will be determined through training process. Firstly, we use the ``Xavier'' initialization \cite{glorot2010understanding} to assign initial random values to these to be determined CNN parameters. This initializer is designed to keep the scale of gradients roughly the same in all layers. Then we use the binary output scores $s$ from our network to calculate the confidence for a GW signal by sigmoid function:
\begin{equation}\label{eq:prob}
p = \frac{1}{1+e^{-s}} \,.
\end{equation}
Then we use binary cross-entropy loss function to evaluate deviation between the predicted values and the actual values in the training data. Based on this estimation mini-batch Adam algorithms \cite{Kingma:2014vow} is applied to optimize the kernel entries in CNN. Here we caution that this confidence value can not be interpreted as the statistical significance of a detection \cite{Gebhard:2019ldz}.

Within every training epoch (i.e., a full pass over all training data) not only the entire training/test data set is randomly shuffled, but also the background noises are newly  sampled in random manner from O1 data except the three GW events. At the end of every epoch, the performance of the network during the training is evaluated by average accuracy for the networks on each mini-batch. We set the learning rate 0.003 and batch size 16. During the curriculum learning, we gradually decrease the signal SNR of both training data and test data from 1 to 0.02. The training process is done within 6 hours on 4 NVIDIA GeForce GTX 1080Ti GPUs, each with 11GB of memory. All the implementations of current work were coded with Python based on the MXNet framework \cite{2015arXiv151201274C}.

\subsection{Accuracy and efficiency test of the neural network}
\begin{figure}
\begin{tabular}{c}
\includegraphics[width=0.5\textwidth]{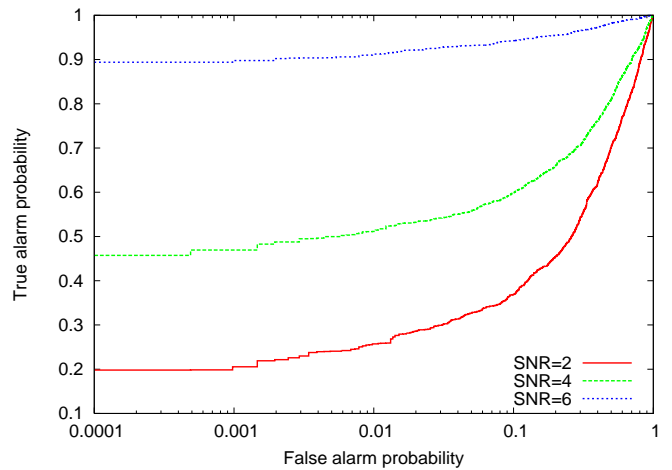}
\end{tabular}
\caption{The ROC curves for test data sets containing signals with matched filtering SNR 2, 4 and 6 respectively. We plot the true alarm probability versus the false alarm probability estimated by our adjusted CNN network.}\label{fig2}
\end{figure}
The authors in \cite{PhysRevLett.120.141103} have compared with the GW signal recognition accuracy and efficiency between CNN network and usual matched filtering technique. Based on the test data set described in the above subsection, we can calculate corresponding true alarm probability and false alarm probability. Then we can construct the receiver operator characteristic (ROC) curves in the Fig.~\ref{fig2}.
\begin{figure}
\begin{tabular}{c}
\includegraphics[width=0.5\textwidth]{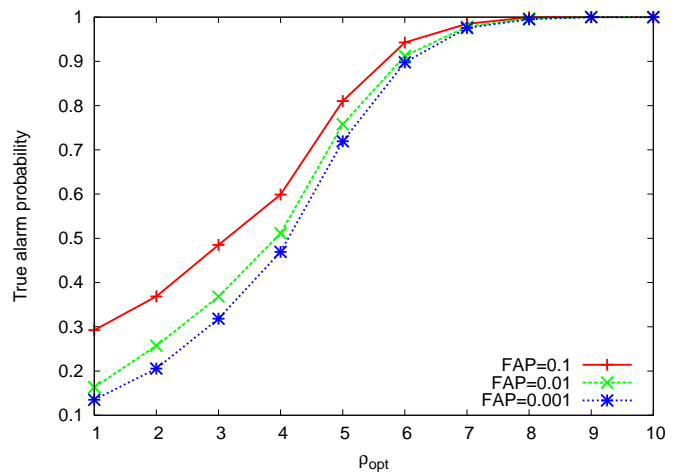}
\end{tabular}
\caption{Efficiency curves for false alarm probabilities 0.1, 0.01 and 0.001 respectively. The true alarm probability is plotted as a function of the optimal SNR.}\label{fig3}
\end{figure}

Stronger signal is easier to be recognized. For a given SNR, the true alarm probability can be used to describe the recognition efficiency. Respect to representative false alarm probability 0.1, 0.01 and 0.001, we plot out the efficiency curves in the Fig.~\ref{fig3}. Both the ROC curves and the efficiency curves are comparable to the results got in \cite{PhysRevLett.120.141103} which means our adjusted neural network is powerful to recognize the GW signal buried in the noise.
\section{Search results of the real LIGO data}\label{sec:result}
\begin{figure*}
\begin{tabular}{ccc}
\includegraphics[width=0.33\textwidth]{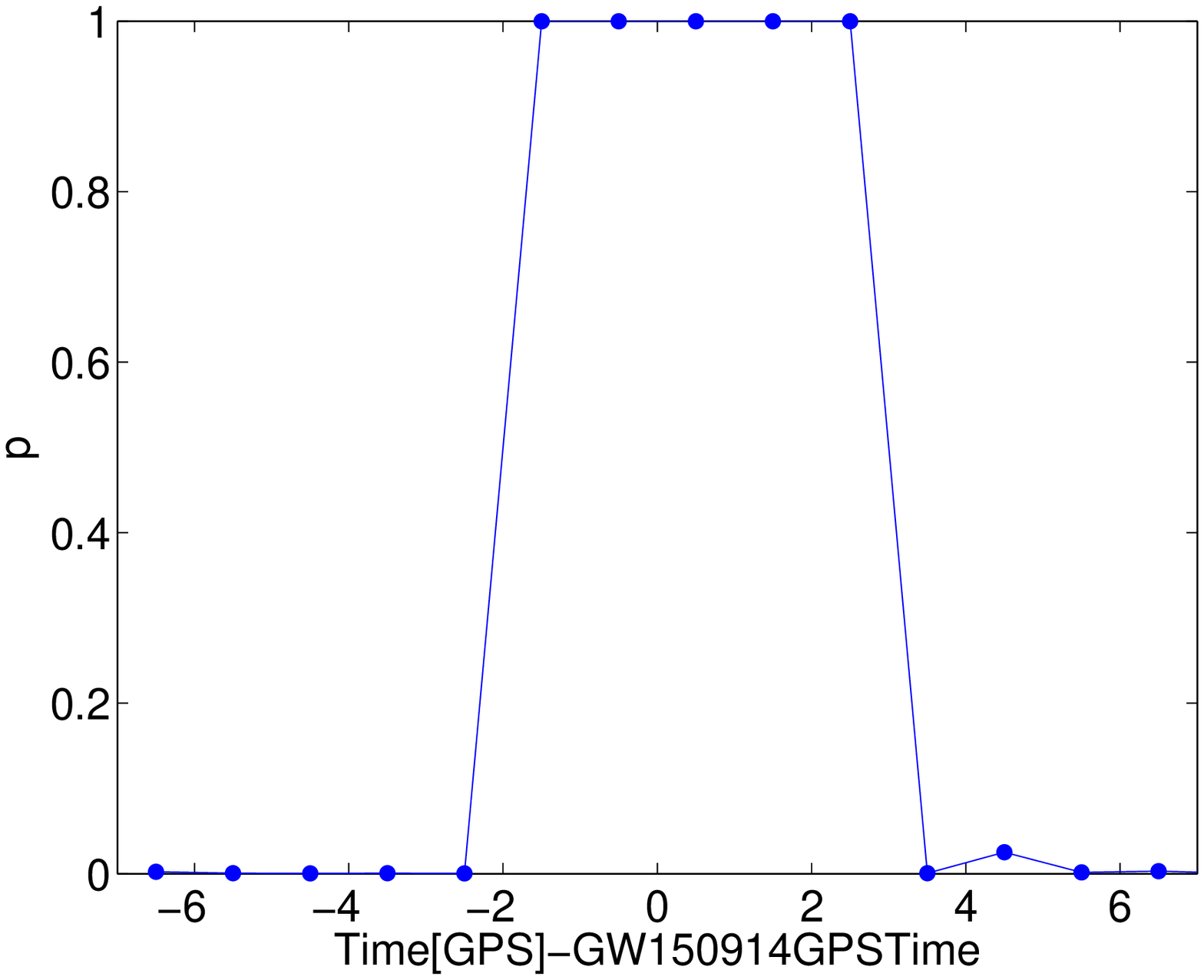}&
\includegraphics[width=0.33\textwidth]{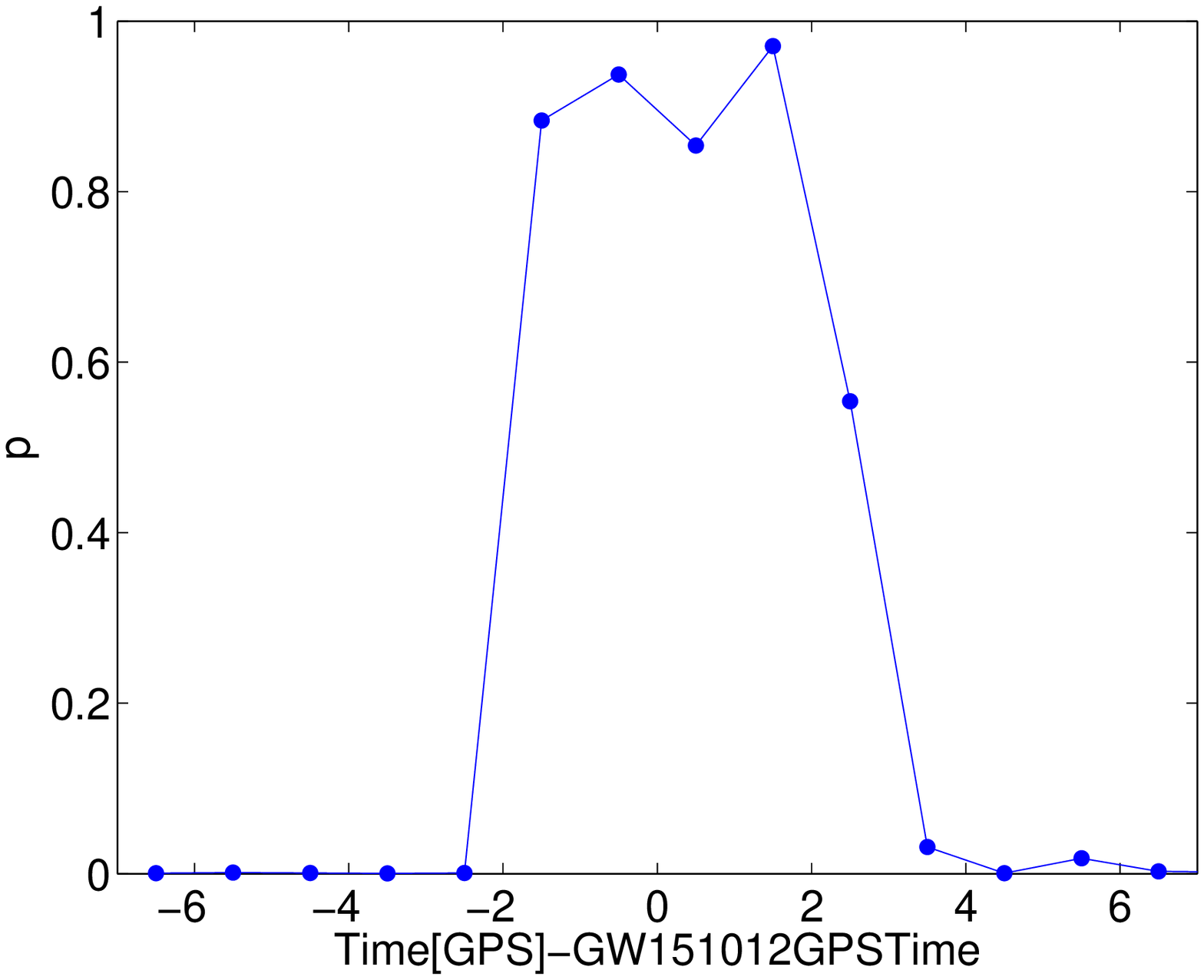}&
\includegraphics[width=0.33\textwidth]{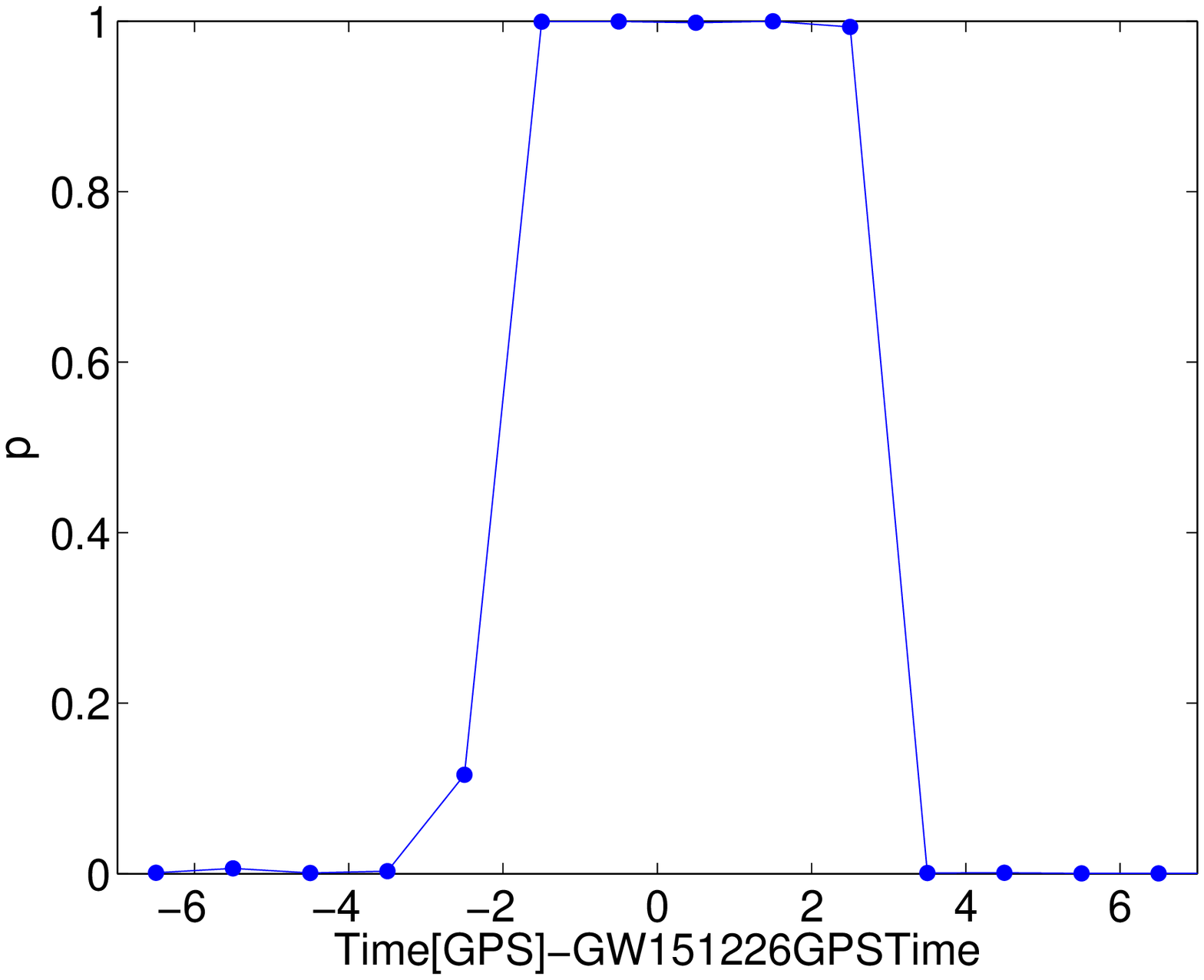}
\end{tabular}
\caption{The output confidence values of our adjusted CNN near the three GW events of O1.}\label{fig4}
\end{figure*}
\begin{figure*}
\begin{tabular}{cccc}
\includegraphics[width=0.25\textwidth]{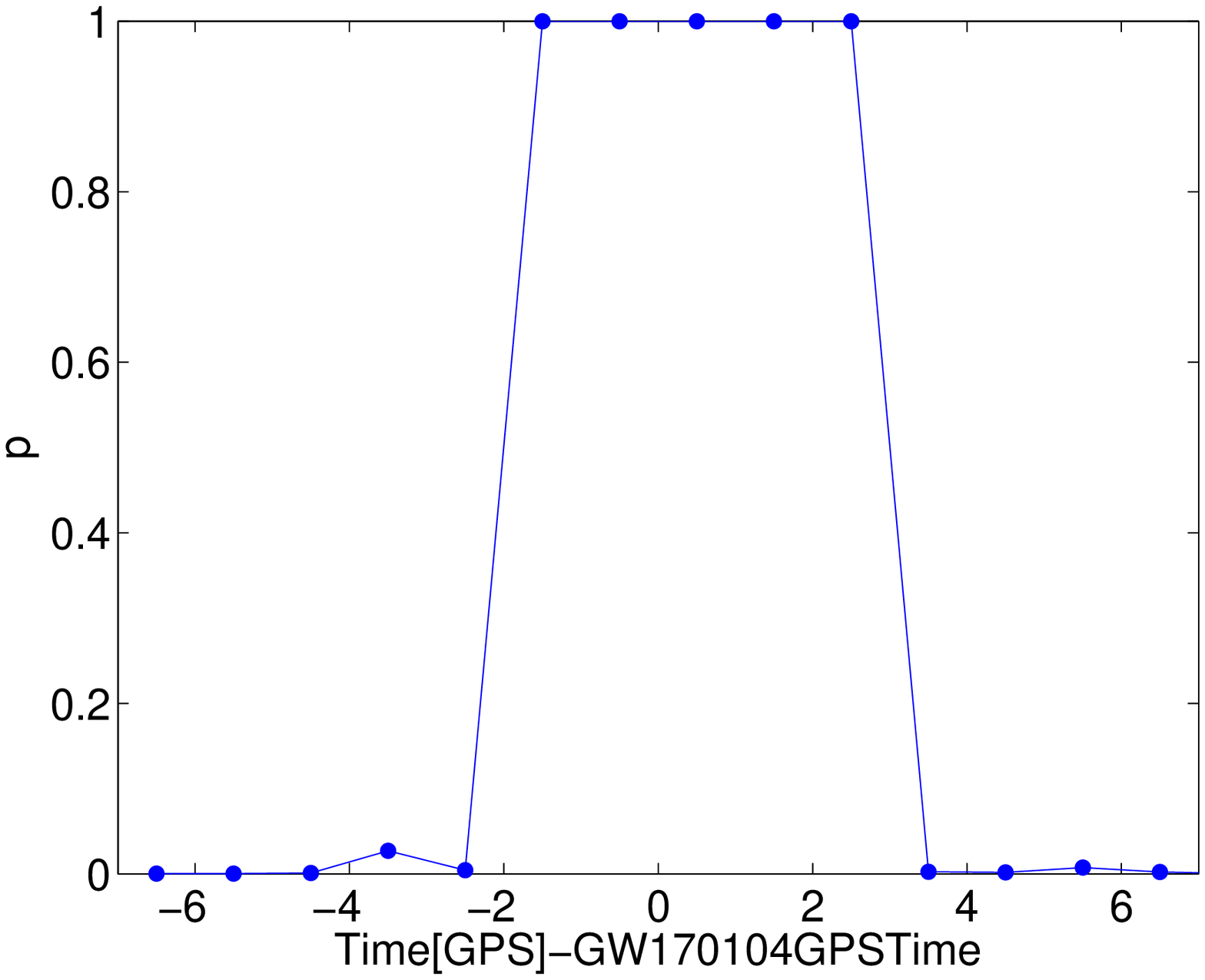}&
\includegraphics[width=0.25\textwidth]{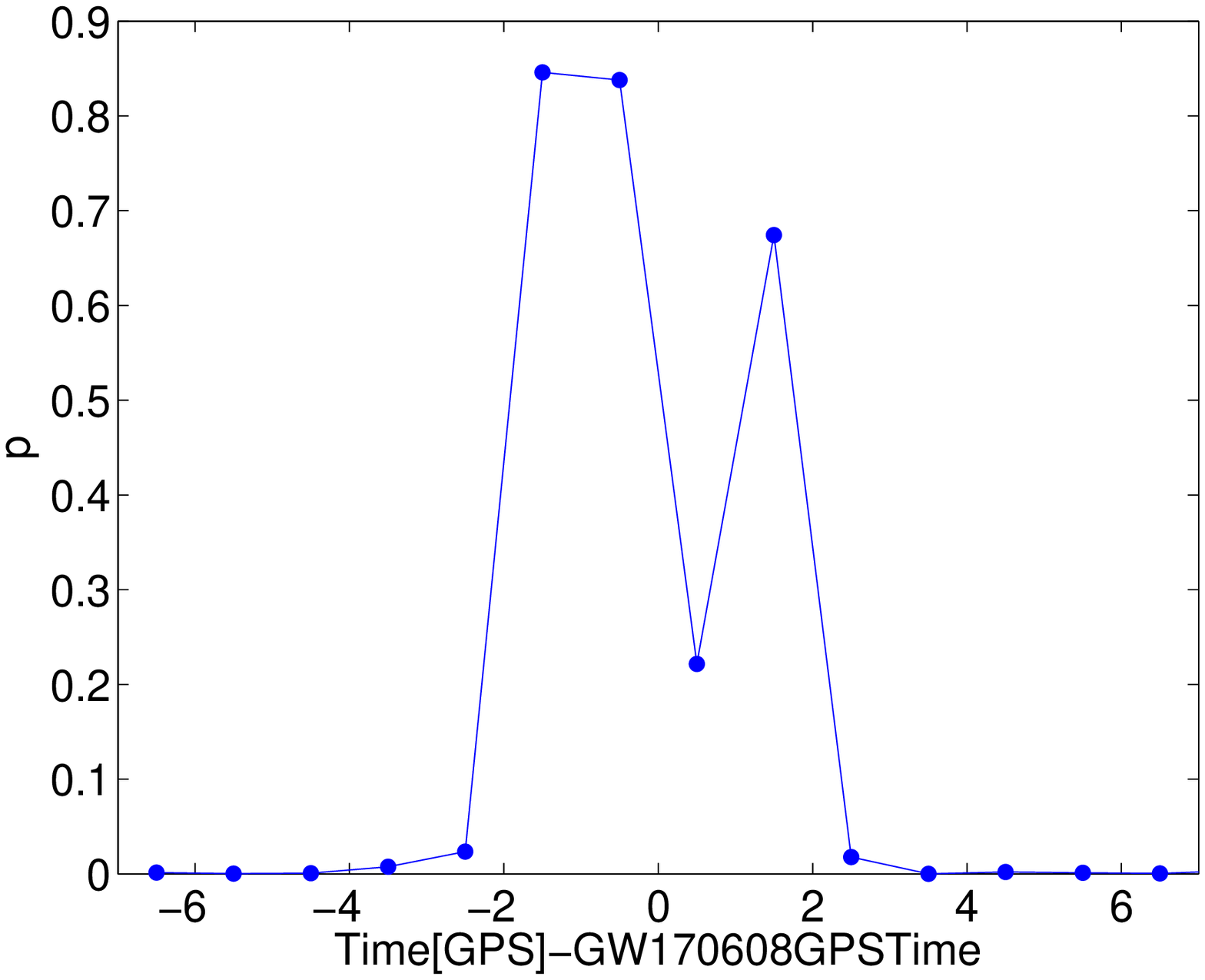}&
\includegraphics[width=0.25\textwidth]{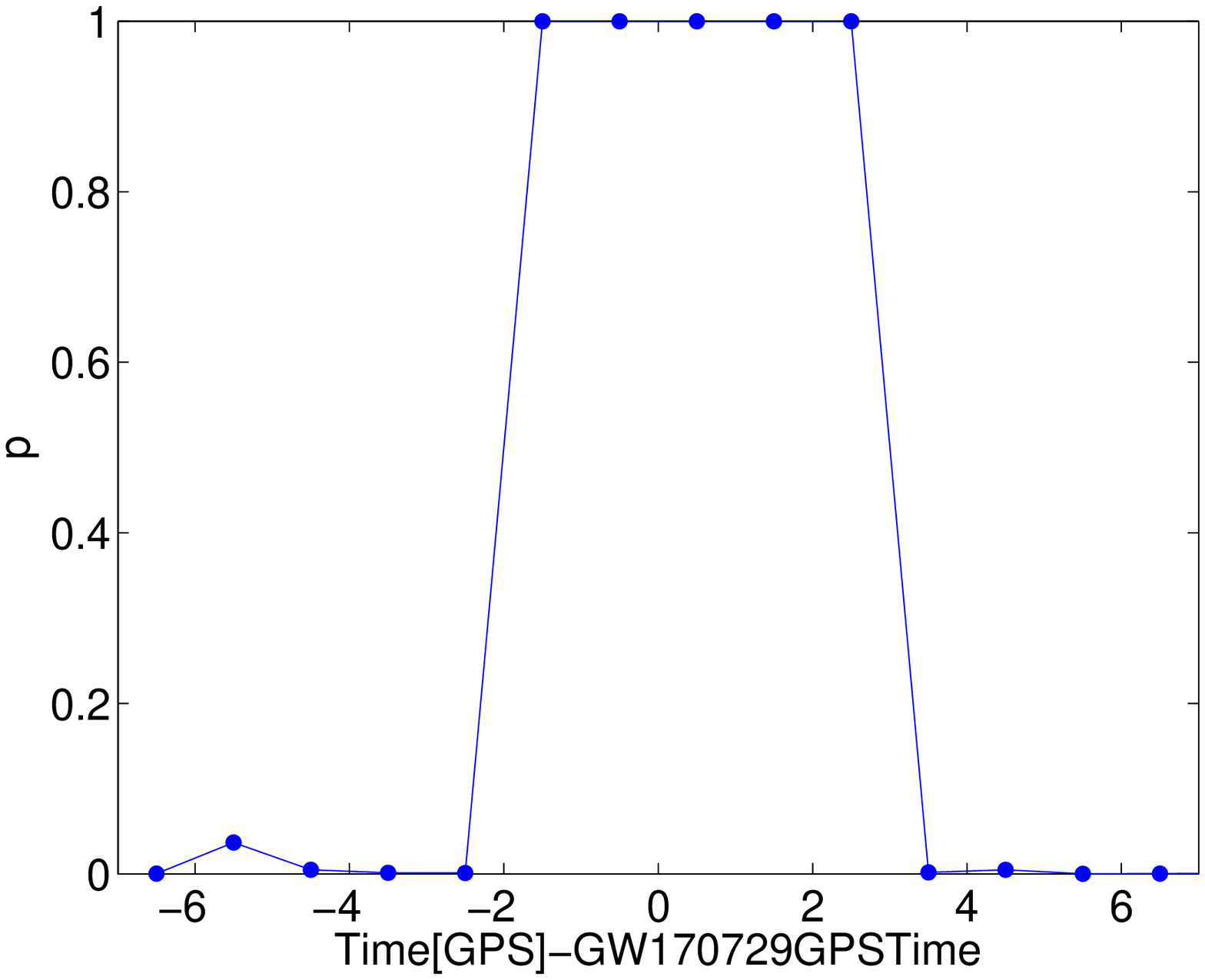}&
\includegraphics[width=0.25\textwidth]{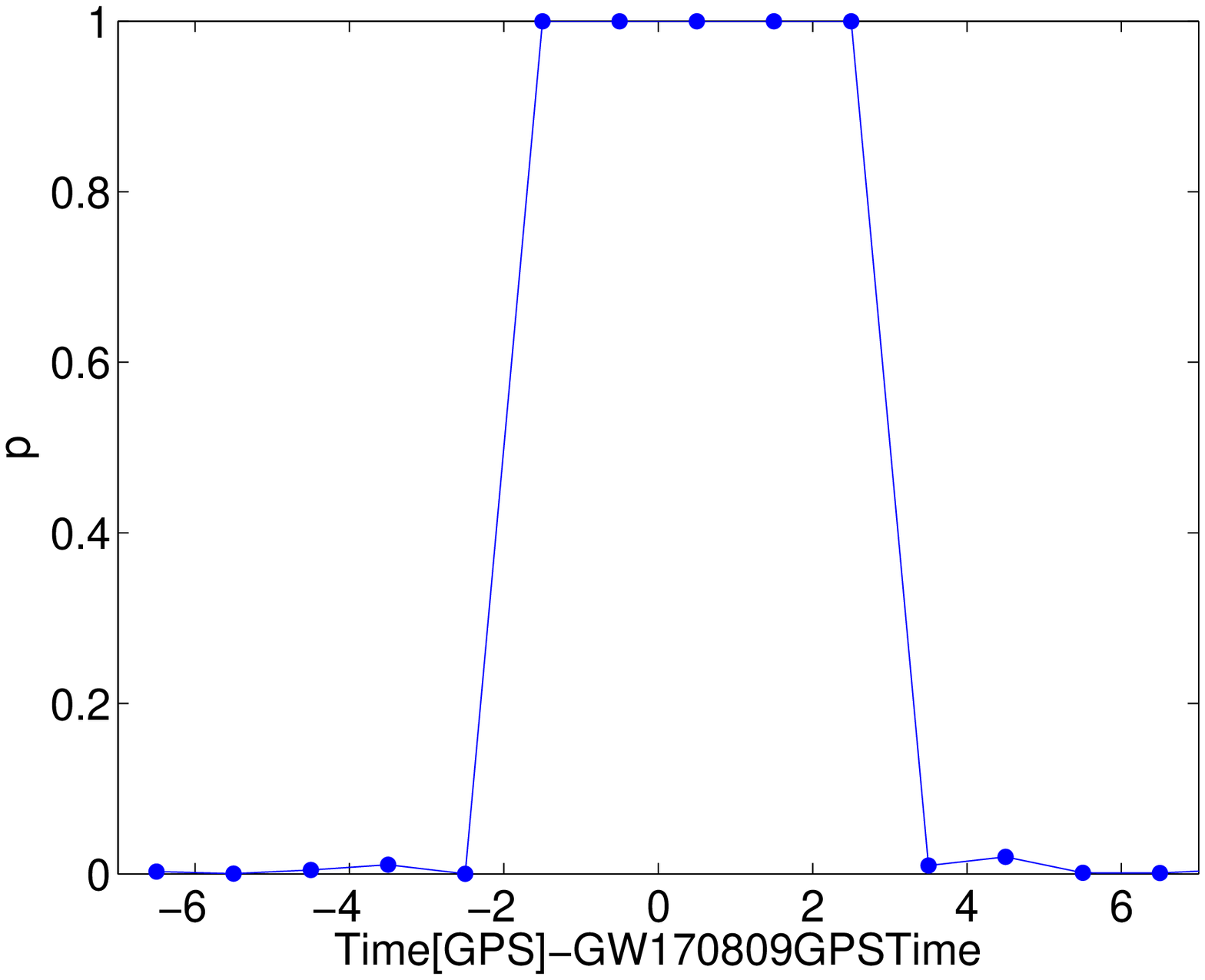}\\
\includegraphics[width=0.25\textwidth]{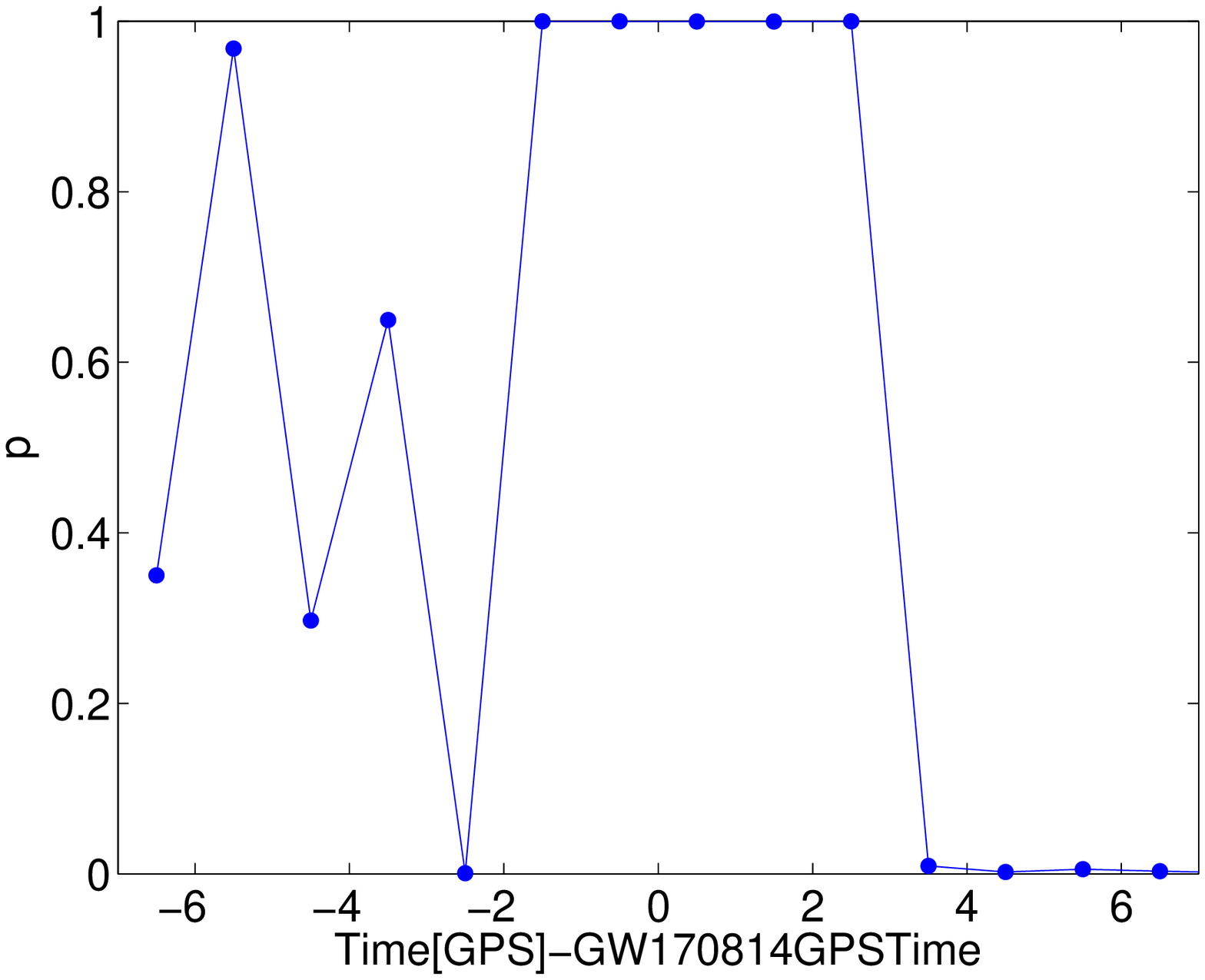}&
\includegraphics[width=0.25\textwidth]{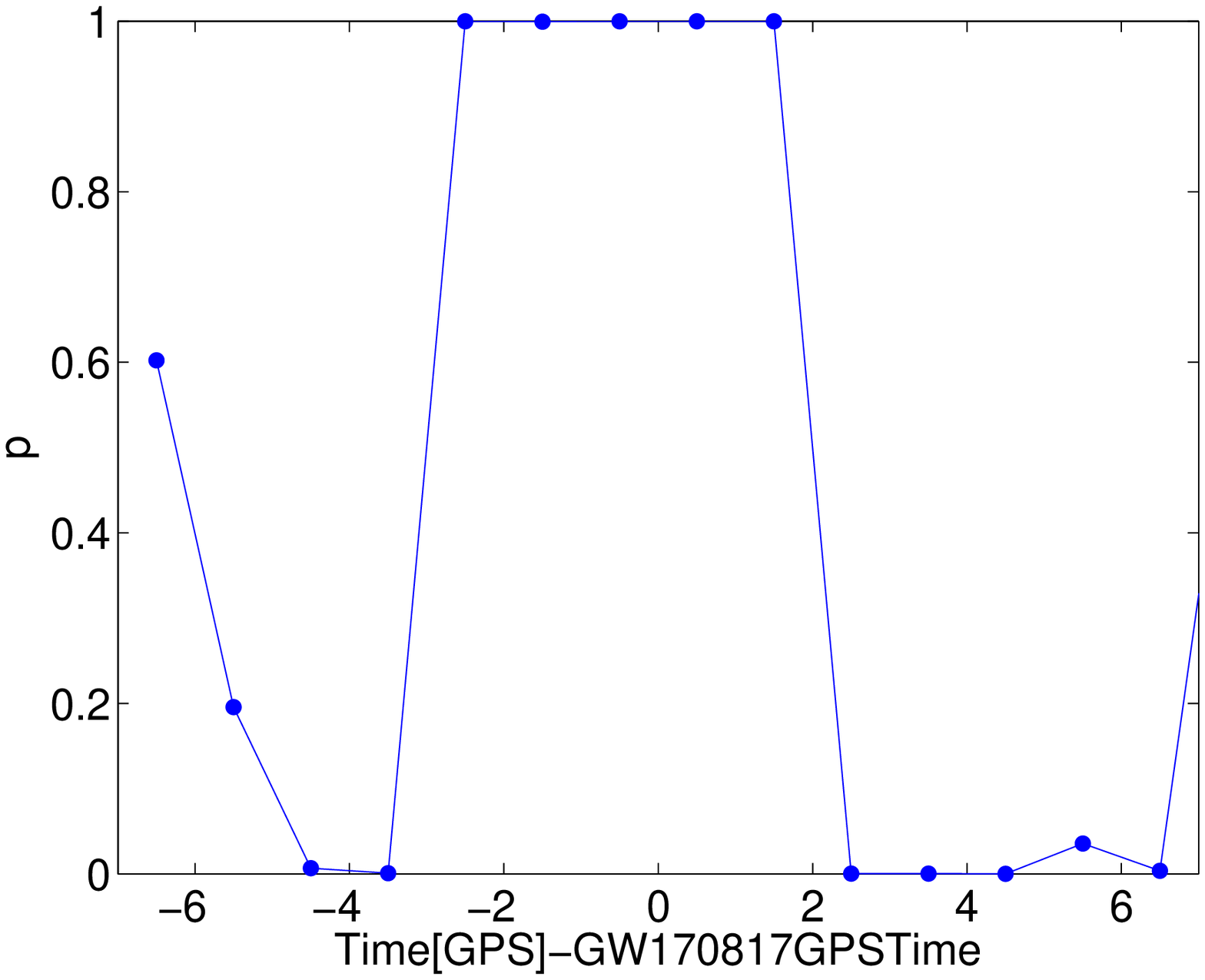}&
\includegraphics[width=0.25\textwidth]{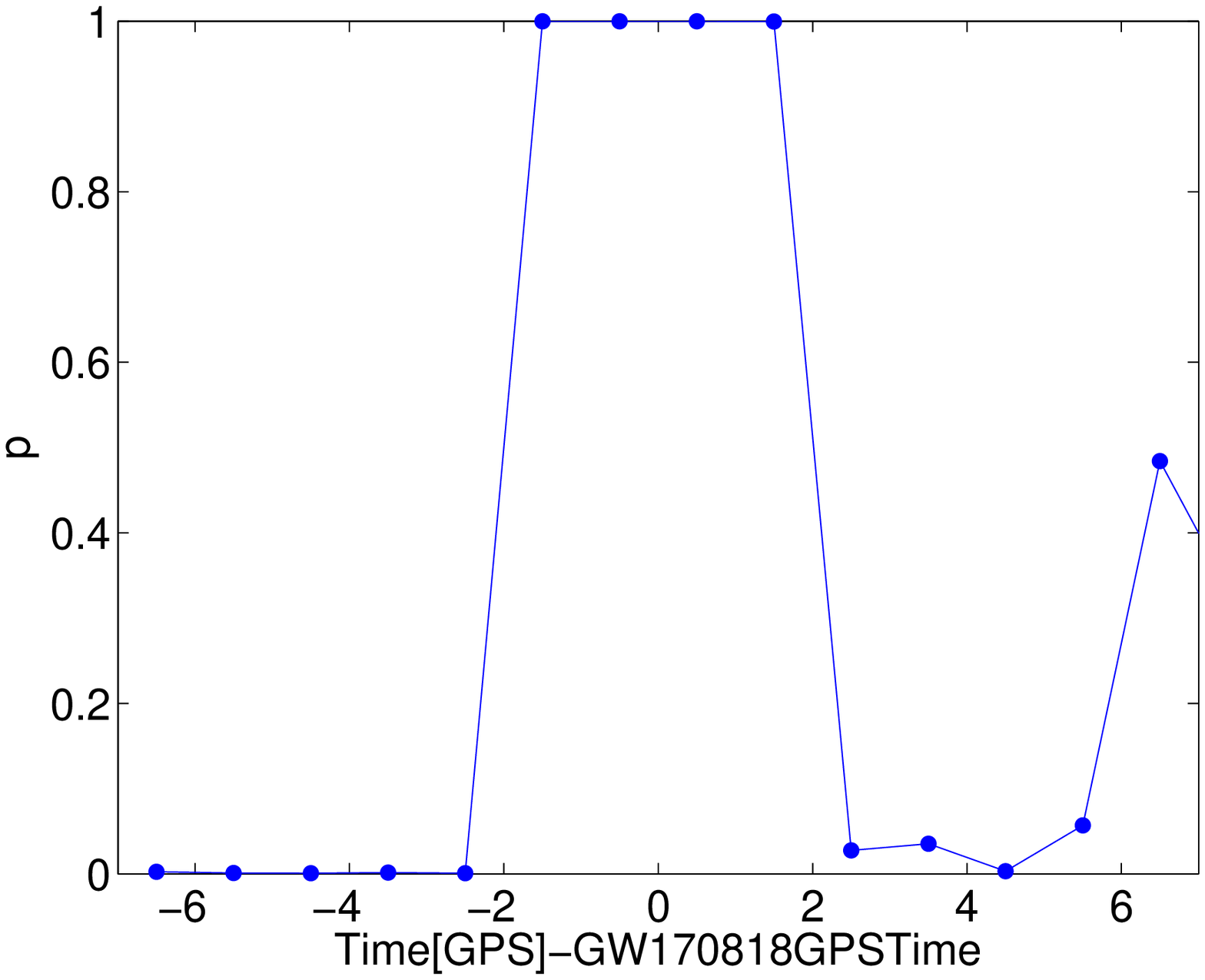}&
\includegraphics[width=0.25\textwidth]{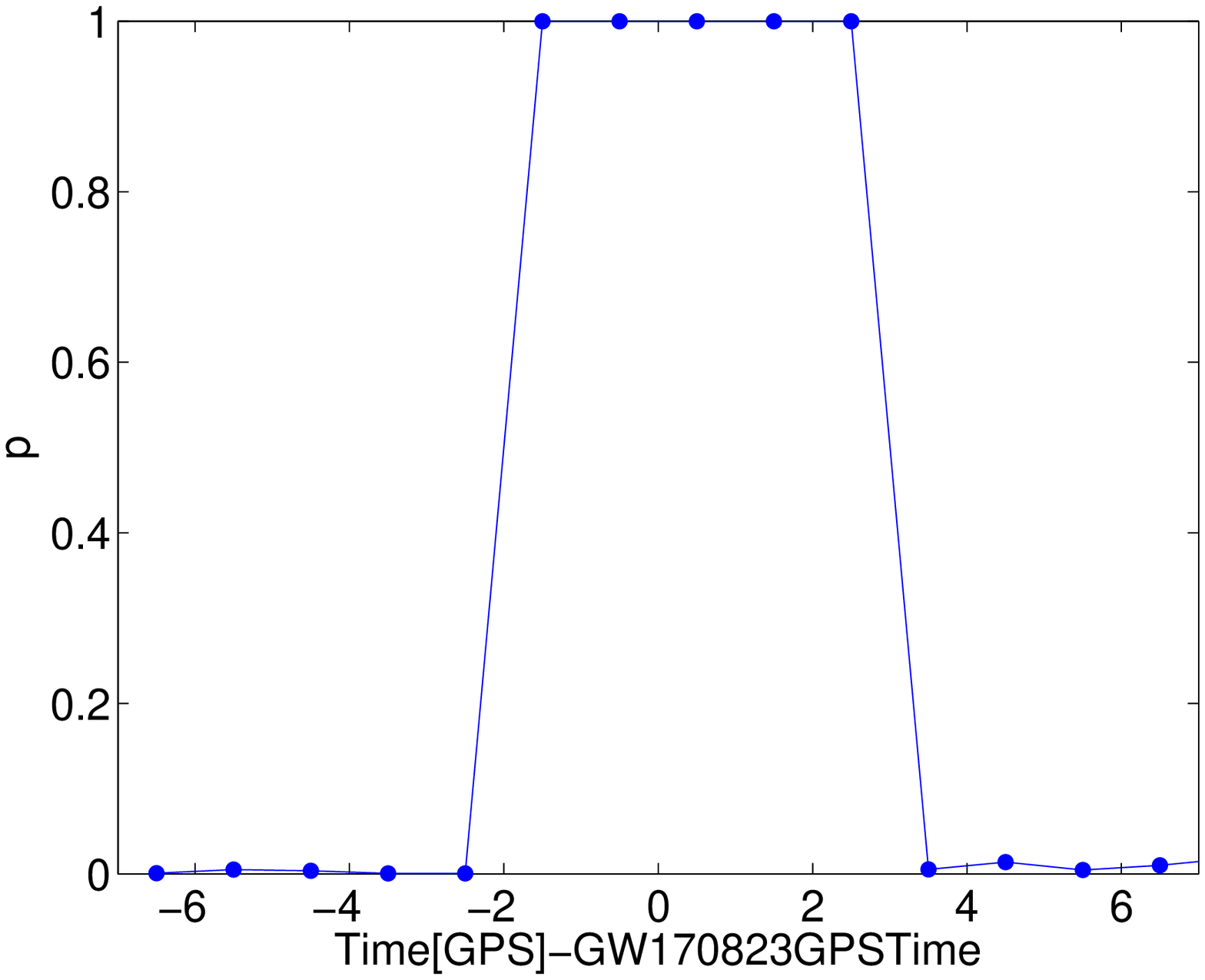}
\end{tabular}
\caption{The output confidence values of our adjusted CNN trained with O1 noise near the eight GW events of O2.}\label{fig5}
\end{figure*}
After being trained, our adjusted CNN is ready to trigger GW signal for O1 data. Each time we take a 5-seconds long data segment from O1 data. After one segment, we move forward one second to get another 5-seconds long data segment. So each one second long data will appear in 5 segments and be processed 5 times by our network. If we assume our network can recognize GW signal within one second time duration, our network will alert continuously 5 times for a true signal.

If the output confidence value is bigger than a given threshold value $p>p_c$, our network gives an alert. If more than 5 continuous alerts happen, a trigger for GW signal will be given.

In the Fig.~\ref{fig4} we plot the confidence values outputted by our adjusted CNN near the GW150914, GW151012 and GW151226 events for O1. We can see our network can mark out all of these three events clearly. For GW150914 and GW151226 there are 5 continuous confidence values approaching 1 while other confidence values are much smaller. For GW151012 the confidence values marking out the signal are not as big as GW150914 and GW151226, but still bigger than 0.5 which is also clearly bigger than the nearby values.

It is interesting to check the effect of the noise used in the training data set and the real data. The authors in \cite{Gebhard:2019ldz} used the network trained with O1 noise to treat the O2 data. We do the same test here. We apply the trained network described in the above section directly to O2 data. We show the confidence values for the eight GW events for O2 in the Fig.~\ref{fig5}. All of the eight events have been clearly marked out. The authors in \cite{Gebhard:2019ldz} shows that the network trained with binary black hole coalescence signal may fail to find out the binary neutron star inspiral signal. This is not true to our adjusted network. We can mark out the GW170817 event very clearly.

Besides the three confirmed GW events, we have also applied our network to all of the O1 data. In the current work we set $p_c=0.5$. If there are more than 5 continuous $p>p_c$, we output a signal alarm at the time of these first $p>p_c$.

If leaving the data quality alone, we find 3955 triggers in O1 which include GW150914, GW151012 and GW151226. If we consider only the data which pass the CBC-CAT3 test there are 2242 triggers in O1. Averagely there are one trigger about every one and half hours. As noted by \cite{Gebhard:2019ldz}, deep learning can not assign a significance to each trigger. So we can not tell which triggers are more believable than others. We can only suggest these triggers are deserved to be checked in more detail with other analysis methods. We have listed these triggered data with center time and time duration on Github~\footnote{\url{https://github.com/WuShichao/mfcnn_catalog}}.

The authors of GWTC-1 reported 3 GW events candidates besides the three known GW events in O1 in the table II of \cite{LIGOScientific:2018mvr}. All these 3 candidates are different to our triggers.

The authors of \cite{Nitz_2019} reported the first open gravitational wave catalog (1-OGC), where 20 GW event candidates are reported including the three known GW events. All these 17 subthreshold candidates are also different to our triggers.

We have also checked the consistency between our triggers and the gamma ray burst (GRB) events listed in \cite{GRBGCN}. There are 1209 GRB events recorded during the O1 run of LIGO. But there is no consistent events found between our triggers and these GRB records.
\section{\label{sec:Summary}Summary}
There are many works published about applying deep learning technique to gravitational wave data analysis in the past few years. Most of these works used simulated data. In the current paper we designed an adjusted CNN and applied it to the whole O1 data of LIGO.

Being trained with noise taken from O1 data and simulated binary black hole coalescence waveform our network can find out the three confirmed GW events clearly. In addition, we used this trained network directly to the 8 GW events found in O2 and we can also mark out all these 8 events clearly. Although this test is inconsistent due to different noise behavior between O1 and O2, our test results indicate that our network and method are robust respect to the training data set. Similar investigation has been done in \cite{Gebhard:2019ldz}.

Besides the three GW events included in O1 we have also found more than 2000 candidates. As noted by the authors of \cite{Gebhard:2019ldz}, we can not assign a significance to each candidate. So we simply call these candidates triggers. We mean these trigged data segments are deserved to be investigated in detail through other means.

We have compare our more than 2000 triggers with the subthreshold events reported in GWTC-1 \cite{LIGOScientific:2018mvr} and in 1-OGC \cite{Nitz_2019}. No consistent events are found between these subthreshold events and our triggers. And more we have also checked the reported GRB events listed in \cite{GRBGCN} during the O1 observation time of LIGO. We have not found consistent events between these GRB events and our triggers either.
%|--------------------------------------------------------------------|
\acknowledgments
%|--------------------------------------------------------------------|
We are thankful to Hao Wei and Jing Li for many helpful discussions. This work was supported by the NSFC (No.~11690023 and No.~11622546) and by the Collaborative research program of the Institute for Cosmic Ray Research (ICRR), the University of Tokyo. Z. Cao was supported by ``the Fundamental Research Funds for the Central Universities", ``the Interdiscipline Research Funds of Beijing Normal University" and the Strategic Priority Research Program of the Chinese Academy of Sciences, grant No. XDB23040100.
%|--------------------------------------------------------------------|

\appendix
\section{Convolution operation in the first layer of the adjusted CNN network}\label{app}
For a template $h(t)$ and a detected strain data $d(t)$, the matched-filter SNR is defined as
\begin{align}
&\rho^2(t_c)\equiv\frac{\left|\langle d|h \rangle(t_c)\right|^2}{\langle h|h \rangle},\\
&\langle d|h \rangle (t_c) = 4\int^\infty_0\frac{\tilde{d}(f)\tilde{h}^*(f)}{S_n(f)}e^{2\pi ift_c}df,\label{eq:corr}\\
&\langle h|h \rangle = 4\int^\infty_0\frac{\tilde{h}(f)\tilde{h}^*(f)}{S_n(f)}df,
\end{align}
where $|\cdot|$ means taking the absolute value, $*$ means complex conjugate and $\tilde{}$ is the Fourier transformation like
\begin{equation}
\tilde{d}(f)=\int^\infty_{-\infty}d(t)e^{-2\pi ift}dt \,.
\end{equation}
The quantity $S_n(f)$ is the one-sided average power spectral density of the detector noise.
Due to the convolution theorem and the relationship between convolution and correlation, the correlation in Eq.~(\ref{eq:corr}) can be rewritten as
\begin{align}
\langle d|h \rangle (t) &= 4\int^\infty_0\frac{\tilde{d}(f)\cdot\tilde{h}^*(f)}{S_n(f)}e^{2\pi ift}df \nonumber\\
 &= 4\int^\infty_0[\tilde{d}(f)\bar{S}_n(f)]\cdot[\tilde{h}(f)\bar{S}_n(f)]^*e^{2\pi ift}df \nonumber\\
 &= 4\int^\infty_0\tilde{\bar{d}}(f)\cdot\tilde{\bar{h}}^*(f)e^{2\pi ift}df \nonumber\\
 &= 2 \,\bar{d}(t)\ast\bar{h}(-t) \,,\\
\bar{d}(t) &= d(t) \ast \bar{S}_n(t) \,, \\
\bar{h}(t) &= h(t) \ast \bar{S}_n(t) \,, \\
\bar{S}_n(t)&=\int^{+\infty}_{-\infty}S_n^{-1/2}(f)e^{2\pi ift}df \,.
\end{align}
where $d(t)\ast h(t)$ means the convolution of functions $d(t)$ and $h(t)$. Similarly, the $\langle h|h \rangle$ can also be calculated in this way as $\langle h|h \rangle = 2 \left.[\bar{h}(t) \ast \bar{h}(-t)]\right|_{t=0}$.

\bibliographystyle{apsrev}
\bibliography{refs}

\begin{thebibliography}{58}
\expandafter\ifx\csname natexlab\endcsname\relax\def\natexlab#1{#1}\fi
\expandafter\ifx\csname bibnamefont\endcsname\relax
  \def\bibnamefont#1{#1}\fi
\expandafter\ifx\csname bibfnamefont\endcsname\relax
  \def\bibfnamefont#1{#1}\fi
\expandafter\ifx\csname citenamefont\endcsname\relax
  \def\citenamefont#1{#1}\fi
\expandafter\ifx\csname url\endcsname\relax
  \def\url#1{\texttt{#1}}\fi
\expandafter\ifx\csname urlprefix\endcsname\relax\def\urlprefix{URL }\fi
\providecommand{\bibinfo}[2]{#2}
\providecommand{\eprint}[2][]{\url{#2}}

\bibitem[{\citenamefont{Abbott
  et~al.}(2016{\natexlab{a}})}]{PhysRevLett.116.061102}
\bibinfo{author}{\bibfnamefont{B.~P.} \bibnamefont{Abbott}}
  \bibnamefont{et~al.} (\bibinfo{collaboration}{LIGO Scientific Collaboration
  and Virgo Collaboration}), \bibinfo{journal}{Phys. Rev. Lett.}
  \textbf{\bibinfo{volume}{116}}, \bibinfo{pages}{061102}
  (\bibinfo{year}{2016}{\natexlab{a}}),
  \urlprefix\url{https://link.aps.org/doi/10.1103/PhysRevLett.116.061102}.

\bibitem[{\citenamefont{Abbott
  et~al.}(2016{\natexlab{b}})}]{PhysRevLett.116.241103}
\bibinfo{author}{\bibfnamefont{B.~P.} \bibnamefont{Abbott}}
  \bibnamefont{et~al.} (\bibinfo{collaboration}{LIGO Scientific Collaboration
  and Virgo Collaboration}), \bibinfo{journal}{Phys. Rev. Lett.}
  \textbf{\bibinfo{volume}{116}}, \bibinfo{pages}{241103}
  (\bibinfo{year}{2016}{\natexlab{b}}),
  \urlprefix\url{https://link.aps.org/doi/10.1103/PhysRevLett.116.241103}.

\bibitem[{\citenamefont{Abbott et~al.}(2016{\natexlab{c}})}]{PhysRevX.6.041015}
\bibinfo{author}{\bibfnamefont{B.~P.} \bibnamefont{Abbott}}
  \bibnamefont{et~al.} (\bibinfo{collaboration}{LIGO Scientific Collaboration
  and Virgo Collaboration}), \bibinfo{journal}{Phys. Rev. X}
  \textbf{\bibinfo{volume}{6}}, \bibinfo{pages}{041015}
  (\bibinfo{year}{2016}{\natexlab{c}}),
  \urlprefix\url{https://link.aps.org/doi/10.1103/PhysRevX.6.041015}.

\bibitem[{\citenamefont{Abbott
  et~al.}(2017{\natexlab{a}})}]{PhysRevLett.118.221101}
\bibinfo{author}{\bibfnamefont{B.~P.} \bibnamefont{Abbott}}
  \bibnamefont{et~al.} (\bibinfo{collaboration}{LIGO Scientific and Virgo
  Collaboration}), \bibinfo{journal}{Phys. Rev. Lett.}
  \textbf{\bibinfo{volume}{118}}, \bibinfo{pages}{221101}
  (\bibinfo{year}{2017}{\natexlab{a}}),
  \urlprefix\url{https://link.aps.org/doi/10.1103/PhysRevLett.118.221101}.

\bibitem[{\citenamefont{Abbott
  et~al.}(2017{\natexlab{b}})}]{PhysRevLett.119.141101}
\bibinfo{author}{\bibfnamefont{B.~P.} \bibnamefont{Abbott}}
  \bibnamefont{et~al.} (\bibinfo{collaboration}{LIGO Scientific Collaboration
  and Virgo Collaboration}), \bibinfo{journal}{Phys. Rev. Lett.}
  \textbf{\bibinfo{volume}{119}}, \bibinfo{pages}{141101}
  (\bibinfo{year}{2017}{\natexlab{b}}),
  \urlprefix\url{https://link.aps.org/doi/10.1103/PhysRevLett.119.141101}.

\bibitem[{\citenamefont{Abbott
  et~al.}(2017{\natexlab{c}})}]{PhysRevLett.119.161101}
\bibinfo{author}{\bibfnamefont{B.~P.} \bibnamefont{Abbott}}
  \bibnamefont{et~al.} (\bibinfo{collaboration}{LIGO Scientific Collaboration
  and Virgo Collaboration}), \bibinfo{journal}{Phys. Rev. Lett.}
  \textbf{\bibinfo{volume}{119}}, \bibinfo{pages}{161101}
  (\bibinfo{year}{2017}{\natexlab{c}}),
  \urlprefix\url{https://link.aps.org/doi/10.1103/PhysRevLett.119.161101}.

\bibitem[{\citenamefont{Abbott
  et~al.}(2017{\natexlab{d}})}]{2041-8205-851-2-L35}
\bibinfo{author}{\bibfnamefont{B.~P.} \bibnamefont{Abbott}}
  \bibnamefont{et~al.}, \bibinfo{journal}{The Astrophysical Journal Letters}
  \textbf{\bibinfo{volume}{851}}, \bibinfo{pages}{L35}
  (\bibinfo{year}{2017}{\natexlab{d}}),
  \urlprefix\url{http://stacks.iop.org/2041-8205/851/i=2/a=L35}.

\bibitem[{\citenamefont{Abbott et~al.}(2018)}]{LIGOScientific:2018mvr}
\bibinfo{author}{\bibfnamefont{B.~P.} \bibnamefont{Abbott}}
  \bibnamefont{et~al.} (\bibinfo{collaboration}{LIGO Scientific, Virgo})
  (\bibinfo{year}{2018}), \eprint{1811.12907}.

\bibitem[{\citenamefont{George and
  Huerta}(2018{\natexlab{a}})}]{PhysRevD.97.044039}
\bibinfo{author}{\bibfnamefont{D.}~\bibnamefont{George}} \bibnamefont{and}
  \bibinfo{author}{\bibfnamefont{E.~A.} \bibnamefont{Huerta}},
  \bibinfo{journal}{Phys. Rev. D} \textbf{\bibinfo{volume}{97}},
  \bibinfo{pages}{044039} (\bibinfo{year}{2018}{\natexlab{a}}),
  \urlprefix\url{https://link.aps.org/doi/10.1103/PhysRevD.97.044039}.

\bibitem[{\citenamefont{George and Huerta}(2018{\natexlab{b}})}]{GEORGE201864}
\bibinfo{author}{\bibfnamefont{D.}~\bibnamefont{George}} \bibnamefont{and}
  \bibinfo{author}{\bibfnamefont{E.}~\bibnamefont{Huerta}},
  \bibinfo{journal}{Physics Letters B} \textbf{\bibinfo{volume}{778}},
  \bibinfo{pages}{64 } (\bibinfo{year}{2018}{\natexlab{b}}), ISSN
  \bibinfo{issn}{0370-2693},
  \urlprefix\url{http://www.sciencedirect.com/science/article/pii/S0370269317310390}.

\bibitem[{\citenamefont{{Li} et~al.}(2017)\citenamefont{{Li}, {Yu}, and
  {Fan}}}]{2017arXiv171200356L}
\bibinfo{author}{\bibfnamefont{X.}~\bibnamefont{{Li}}},
  \bibinfo{author}{\bibfnamefont{W.}~\bibnamefont{{Yu}}}, \bibnamefont{and}
  \bibinfo{author}{\bibfnamefont{X.}~\bibnamefont{{Fan}}},
  \bibinfo{journal}{ArXiv e-prints}  (\bibinfo{year}{2017}),
  \eprint{1712.00356}.

\bibitem[{\citenamefont{Fan et~al.}(2018)\citenamefont{Fan, Li, Li, Zhong, and
  Cao}}]{fanli18}
\bibinfo{author}{\bibfnamefont{X.-L.} \bibnamefont{Fan}},
  \bibinfo{author}{\bibfnamefont{J.}~\bibnamefont{Li}},
  \bibinfo{author}{\bibfnamefont{X.}~\bibnamefont{Li}},
  \bibinfo{author}{\bibfnamefont{Y.}~\bibnamefont{Zhong}}, \bibnamefont{and}
  \bibinfo{author}{\bibfnamefont{J.}~\bibnamefont{Cao}},
  \bibinfo{journal}{SCIENCE CHINA Physics, Mechanics \& Astronomy} p.
  \bibinfo{pages}{in press} (\bibinfo{year}{2018}).

\bibitem[{\citenamefont{Cohen et~al.}(2018)\citenamefont{Cohen, Freytsis, and
  Ostdiek}}]{Cohen2018}
\bibinfo{author}{\bibfnamefont{T.}~\bibnamefont{Cohen}},
  \bibinfo{author}{\bibfnamefont{M.}~\bibnamefont{Freytsis}}, \bibnamefont{and}
  \bibinfo{author}{\bibfnamefont{B.}~\bibnamefont{Ostdiek}},
  \bibinfo{journal}{Journal of High Energy Physics}
  \textbf{\bibinfo{volume}{2018}}, \bibinfo{pages}{34} (\bibinfo{year}{2018}),
  ISSN \bibinfo{issn}{1029-8479},
  \urlprefix\url{https://doi.org/10.1007/JHEP02(2018)034}.

\bibitem[{\citenamefont{Chang et~al.}(2018)\citenamefont{Chang, Cohen, and
  Ostdiek}}]{PhysRevD.97.056009}
\bibinfo{author}{\bibfnamefont{S.}~\bibnamefont{Chang}},
  \bibinfo{author}{\bibfnamefont{T.}~\bibnamefont{Cohen}}, \bibnamefont{and}
  \bibinfo{author}{\bibfnamefont{B.}~\bibnamefont{Ostdiek}},
  \bibinfo{journal}{Phys. Rev. D} \textbf{\bibinfo{volume}{97}},
  \bibinfo{pages}{056009} (\bibinfo{year}{2018}),
  \urlprefix\url{https://link.aps.org/doi/10.1103/PhysRevD.97.056009}.

\bibitem[{\citenamefont{Caron et~al.}(2018)\citenamefont{Caron, Gomez-Vargas,
  Hendriks, and de~Austri}}]{1475-7516-2018-05-058}
\bibinfo{author}{\bibfnamefont{S.}~\bibnamefont{Caron}},
  \bibinfo{author}{\bibfnamefont{G.~A.} \bibnamefont{Gomez-Vargas}},
  \bibinfo{author}{\bibfnamefont{L.}~\bibnamefont{Hendriks}}, \bibnamefont{and}
  \bibinfo{author}{\bibfnamefont{R.~R.} \bibnamefont{de~Austri}},
  \bibinfo{journal}{Journal of Cosmology and Astroparticle Physics}
  \textbf{\bibinfo{volume}{2018}}, \bibinfo{pages}{058} (\bibinfo{year}{2018}),
  \urlprefix\url{http://stacks.iop.org/1475-7516/2018/i=05/a=058}.

\bibitem[{\citenamefont{Abraham et~al.}(2019)\citenamefont{Abraham, Mukund,
  Vibhute, Sharma, Iyyani, Bhattacharya, Rao, Vadawale, and
  Bhalerao}}]{abraham2019machine}
\bibinfo{author}{\bibfnamefont{S.}~\bibnamefont{Abraham}},
  \bibinfo{author}{\bibfnamefont{N.}~\bibnamefont{Mukund}},
  \bibinfo{author}{\bibfnamefont{A.}~\bibnamefont{Vibhute}},
  \bibinfo{author}{\bibfnamefont{V.}~\bibnamefont{Sharma}},
  \bibinfo{author}{\bibfnamefont{S.}~\bibnamefont{Iyyani}},
  \bibinfo{author}{\bibfnamefont{D.}~\bibnamefont{Bhattacharya}},
  \bibinfo{author}{\bibfnamefont{A.~R.} \bibnamefont{Rao}},
  \bibinfo{author}{\bibfnamefont{S.}~\bibnamefont{Vadawale}}, \bibnamefont{and}
  \bibinfo{author}{\bibfnamefont{V.}~\bibnamefont{Bhalerao}},
  \emph{\bibinfo{title}{A machine learning approach for grb detection in
  astrosat czti data}} (\bibinfo{year}{2019}), \eprint{1906.09670}.

\bibitem[{\citenamefont{Charnock and Moss}(2017)}]{charnock2017deep}
\bibinfo{author}{\bibfnamefont{T.}~\bibnamefont{Charnock}} \bibnamefont{and}
  \bibinfo{author}{\bibfnamefont{A.}~\bibnamefont{Moss}}, \bibinfo{journal}{The
  Astrophysical Journal Letters} \textbf{\bibinfo{volume}{837}},
  \bibinfo{pages}{L28} (\bibinfo{year}{2017}).

\bibitem[{\citenamefont{Moss}(2018)}]{Moss:2018tug}
\bibinfo{author}{\bibfnamefont{A.}~\bibnamefont{Moss}} (\bibinfo{year}{2018}),
  \eprint{1810.06441}.

\bibitem[{\citenamefont{Gupta et~al.}(2018)\citenamefont{Gupta, Matilla, Hsu,
  and Haiman}}]{gupta2018non}
\bibinfo{author}{\bibfnamefont{A.}~\bibnamefont{Gupta}},
  \bibinfo{author}{\bibfnamefont{J.~M.~Z.} \bibnamefont{Matilla}},
  \bibinfo{author}{\bibfnamefont{D.}~\bibnamefont{Hsu}}, \bibnamefont{and}
  \bibinfo{author}{\bibfnamefont{Z.}~\bibnamefont{Haiman}},
  \bibinfo{journal}{Physical Review D} \textbf{\bibinfo{volume}{97}},
  \bibinfo{pages}{103515} (\bibinfo{year}{2018}).

\bibitem[{\citenamefont{Shirasaki et~al.}(2019)\citenamefont{Shirasaki,
  Yoshida, and Ikeda}}]{Shirasaki:2018thk}
\bibinfo{author}{\bibfnamefont{M.}~\bibnamefont{Shirasaki}},
  \bibinfo{author}{\bibfnamefont{N.}~\bibnamefont{Yoshida}}, \bibnamefont{and}
  \bibinfo{author}{\bibfnamefont{S.}~\bibnamefont{Ikeda}},
  \bibinfo{journal}{Phys. Rev.} \textbf{\bibinfo{volume}{D100}},
  \bibinfo{pages}{043527} (\bibinfo{year}{2019}), \eprint{1812.05781}.

\bibitem[{\citenamefont{Ribli et~al.}(2019)\citenamefont{Ribli, Pataki,
  Matilla, Hsu, Haiman, and Csabai}}]{Ribli:2019wtw}
\bibinfo{author}{\bibfnamefont{D.}~\bibnamefont{Ribli}},
  \bibinfo{author}{\bibfnamefont{B.~r.} \bibnamefont{Pataki}},
  \bibinfo{author}{\bibfnamefont{J.~M.~Z.} \bibnamefont{Matilla}},
  \bibinfo{author}{\bibfnamefont{D.}~\bibnamefont{Hsu}},
  \bibinfo{author}{\bibfnamefont{Z.}~\bibnamefont{Haiman}}, \bibnamefont{and}
  \bibinfo{author}{\bibfnamefont{I.}~\bibnamefont{Csabai}}
  (\bibinfo{year}{2019}), \eprint{1902.03663}.

\bibitem[{\citenamefont{Fluri et~al.}(2019)\citenamefont{Fluri, Kacprzak,
  Lucchi, Refregier, Amara, Hofmann, and Schneider}}]{Fluri:2019qtp}
\bibinfo{author}{\bibfnamefont{J.}~\bibnamefont{Fluri}},
  \bibinfo{author}{\bibfnamefont{T.}~\bibnamefont{Kacprzak}},
  \bibinfo{author}{\bibfnamefont{A.}~\bibnamefont{Lucchi}},
  \bibinfo{author}{\bibfnamefont{A.}~\bibnamefont{Refregier}},
  \bibinfo{author}{\bibfnamefont{A.}~\bibnamefont{Amara}},
  \bibinfo{author}{\bibfnamefont{T.}~\bibnamefont{Hofmann}}, \bibnamefont{and}
  \bibinfo{author}{\bibfnamefont{A.}~\bibnamefont{Schneider}},
  \bibinfo{journal}{Phys. Rev.} \textbf{\bibinfo{volume}{D100}},
  \bibinfo{pages}{063514} (\bibinfo{year}{2019}), \eprint{1906.03156}.

\bibitem[{\citenamefont{Chua et~al.}(2019)\citenamefont{Chua, Galley, and
  Vallisneri}}]{Chua:2018woh}
\bibinfo{author}{\bibfnamefont{A.~J.~K.} \bibnamefont{Chua}},
  \bibinfo{author}{\bibfnamefont{C.~R.} \bibnamefont{Galley}},
  \bibnamefont{and}
  \bibinfo{author}{\bibfnamefont{M.}~\bibnamefont{Vallisneri}},
  \bibinfo{journal}{Phys. Rev. Lett.} \textbf{\bibinfo{volume}{122}},
  \bibinfo{pages}{211101} (\bibinfo{year}{2019}), \eprint{1811.05491}.

\bibitem[{\citenamefont{Rebei et~al.}(2019)\citenamefont{Rebei, Huerta, Wang,
  Habib, Haas, Johnson, and George}}]{Rebei:2018lzh}
\bibinfo{author}{\bibfnamefont{A.}~\bibnamefont{Rebei}},
  \bibinfo{author}{\bibfnamefont{E.~A.} \bibnamefont{Huerta}},
  \bibinfo{author}{\bibfnamefont{S.}~\bibnamefont{Wang}},
  \bibinfo{author}{\bibfnamefont{S.}~\bibnamefont{Habib}},
  \bibinfo{author}{\bibfnamefont{R.}~\bibnamefont{Haas}},
  \bibinfo{author}{\bibfnamefont{D.}~\bibnamefont{Johnson}}, \bibnamefont{and}
  \bibinfo{author}{\bibfnamefont{D.}~\bibnamefont{George}},
  \bibinfo{journal}{Phys. Rev.} \textbf{\bibinfo{volume}{D100}},
  \bibinfo{pages}{044025} (\bibinfo{year}{2019}), \eprint{1807.09787}.

\bibitem[{\citenamefont{Chua and Vallisneri}(2019)}]{Chua:2019wwt}
\bibinfo{author}{\bibfnamefont{A.~J.~K.} \bibnamefont{Chua}} \bibnamefont{and}
  \bibinfo{author}{\bibfnamefont{M.}~\bibnamefont{Vallisneri}}
  (\bibinfo{year}{2019}), \eprint{1909.05966}.

\bibitem[{\citenamefont{Setyawati et~al.}(2019)\citenamefont{Setyawati, Pürrer,
  and Ohme}}]{Setyawati:2019xzw}
\bibinfo{author}{\bibfnamefont{Y.}~\bibnamefont{Setyawati}},
  \bibinfo{author}{\bibfnamefont{M.}~\bibnamefont{Pürrer}}, \bibnamefont{and}
  \bibinfo{author}{\bibfnamefont{F.}~\bibnamefont{Ohme}}
  (\bibinfo{year}{2019}), \eprint{1909.10986}.

\bibitem[{\citenamefont{Rampone et~al.}(2013)\citenamefont{Rampone, Pierro,
  Troiano, and Pinto}}]{RAMPONE:2013oga}
\bibinfo{author}{\bibfnamefont{S.}~\bibnamefont{Rampone}},
  \bibinfo{author}{\bibfnamefont{V.}~\bibnamefont{Pierro}},
  \bibinfo{author}{\bibfnamefont{L.}~\bibnamefont{Troiano}}, \bibnamefont{and}
  \bibinfo{author}{\bibfnamefont{I.~M.} \bibnamefont{Pinto}},
  \bibinfo{journal}{Int. J. Mod. Phys.} \textbf{\bibinfo{volume}{C24}},
  \bibinfo{pages}{1350084} (\bibinfo{year}{2013}), \eprint{1401.5941}.

\bibitem[{\citenamefont{Mukherjee et~al.}(2010)\citenamefont{Mukherjee, Obaid,
  and Matkarimov}}]{Mukherjee:2010zza}
\bibinfo{author}{\bibfnamefont{S.}~\bibnamefont{Mukherjee}},
  \bibinfo{author}{\bibfnamefont{R.}~\bibnamefont{Obaid}}, \bibnamefont{and}
  \bibinfo{author}{\bibfnamefont{B.}~\bibnamefont{Matkarimov}},
  \bibinfo{journal}{J. Phys. Conf. Ser.} \textbf{\bibinfo{volume}{243}},
  \bibinfo{pages}{012006} (\bibinfo{year}{2010}).

\bibitem[{\citenamefont{Powell et~al.}(2015)\citenamefont{Powell, Trifir¨°,
  Cuoco, Heng, and Cavagli¨¤}}]{Powell:2015ona}
\bibinfo{author}{\bibfnamefont{J.}~\bibnamefont{Powell}},
  \bibinfo{author}{\bibfnamefont{D.}~\bibnamefont{Trifir¨°}},
  \bibinfo{author}{\bibfnamefont{E.}~\bibnamefont{Cuoco}},
  \bibinfo{author}{\bibfnamefont{I.~S.} \bibnamefont{Heng}}, \bibnamefont{and}
  \bibinfo{author}{\bibfnamefont{M.}~\bibnamefont{Cavagli¨¤}},
  \bibinfo{journal}{Class. Quant. Grav.} \textbf{\bibinfo{volume}{32}},
  \bibinfo{pages}{215012} (\bibinfo{year}{2015}), \eprint{1505.01299}.

\bibitem[{\citenamefont{Powell et~al.}(2017)\citenamefont{Powell,
  Torres-Forn¨¦, Lynch, Trifir¨°, Cuoco, Cavagli¨¤, Heng, and
  Font}}]{Powell:2016rkl}
\bibinfo{author}{\bibfnamefont{J.}~\bibnamefont{Powell}},
  \bibinfo{author}{\bibfnamefont{A.}~\bibnamefont{Torres-Forn¨¦}},
  \bibinfo{author}{\bibfnamefont{R.}~\bibnamefont{Lynch}},
  \bibinfo{author}{\bibfnamefont{D.}~\bibnamefont{Trifir¨°}},
  \bibinfo{author}{\bibfnamefont{E.}~\bibnamefont{Cuoco}},
  \bibinfo{author}{\bibfnamefont{M.}~\bibnamefont{Cavagli¨¤}},
  \bibinfo{author}{\bibfnamefont{I.~S.} \bibnamefont{Heng}}, \bibnamefont{and}
  \bibinfo{author}{\bibfnamefont{J.~A.} \bibnamefont{Font}},
  \bibinfo{journal}{Class. Quant. Grav.} \textbf{\bibinfo{volume}{34}},
  \bibinfo{pages}{034002} (\bibinfo{year}{2017}), \eprint{1609.06262}.

\bibitem[{\citenamefont{George et~al.}(2018)\citenamefont{George, Shen, and
  Huerta}}]{PhysRevD.97.101501}
\bibinfo{author}{\bibfnamefont{D.}~\bibnamefont{George}},
  \bibinfo{author}{\bibfnamefont{H.}~\bibnamefont{Shen}}, \bibnamefont{and}
  \bibinfo{author}{\bibfnamefont{E.~A.} \bibnamefont{Huerta}},
  \bibinfo{journal}{Phys. Rev. D} \textbf{\bibinfo{volume}{97}},
  \bibinfo{pages}{101501} (\bibinfo{year}{2018}),
  \urlprefix\url{https://link.aps.org/doi/10.1103/PhysRevD.97.101501}.

\bibitem[{\citenamefont{Mukund et~al.}(2017)\citenamefont{Mukund, Abraham,
  Kandhasamy, Mitra, and Philip}}]{PhysRevD.95.104059}
\bibinfo{author}{\bibfnamefont{N.}~\bibnamefont{Mukund}},
  \bibinfo{author}{\bibfnamefont{S.}~\bibnamefont{Abraham}},
  \bibinfo{author}{\bibfnamefont{S.}~\bibnamefont{Kandhasamy}},
  \bibinfo{author}{\bibfnamefont{S.}~\bibnamefont{Mitra}}, \bibnamefont{and}
  \bibinfo{author}{\bibfnamefont{N.~S.} \bibnamefont{Philip}},
  \bibinfo{journal}{Phys. Rev. D} \textbf{\bibinfo{volume}{95}},
  \bibinfo{pages}{104059} (\bibinfo{year}{2017}),
  \urlprefix\url{https://link.aps.org/doi/10.1103/PhysRevD.95.104059}.

\bibitem[{\citenamefont{Razzano and Cuoco}(2018)}]{0264-9381-35-9-095016}
\bibinfo{author}{\bibfnamefont{M.}~\bibnamefont{Razzano}} \bibnamefont{and}
  \bibinfo{author}{\bibfnamefont{E.}~\bibnamefont{Cuoco}},
  \bibinfo{journal}{Classical and Quantum Gravity}
  \textbf{\bibinfo{volume}{35}}, \bibinfo{pages}{095016}
  (\bibinfo{year}{2018}),
  \urlprefix\url{http://stacks.iop.org/0264-9381/35/i=9/a=095016}.

\bibitem[{\citenamefont{{Zevin} et~al.}(2017)\citenamefont{{Zevin}, {Coughlin},
  {Bahaadini}, {Besler}, {Rohani}, {Allen}, {Cabero}, {Crowston},
  {Katsaggelos}, {Larson} et~al.}}]{2017CQGra..34f4003Z}
\bibinfo{author}{\bibfnamefont{M.}~\bibnamefont{{Zevin}}},
  \bibinfo{author}{\bibfnamefont{S.}~\bibnamefont{{Coughlin}}},
  \bibinfo{author}{\bibfnamefont{S.}~\bibnamefont{{Bahaadini}}},
  \bibinfo{author}{\bibfnamefont{E.}~\bibnamefont{{Besler}}},
  \bibinfo{author}{\bibfnamefont{N.}~\bibnamefont{{Rohani}}},
  \bibinfo{author}{\bibfnamefont{S.}~\bibnamefont{{Allen}}},
  \bibinfo{author}{\bibfnamefont{M.}~\bibnamefont{{Cabero}}},
  \bibinfo{author}{\bibfnamefont{K.}~\bibnamefont{{Crowston}}},
  \bibinfo{author}{\bibfnamefont{A.~K.} \bibnamefont{{Katsaggelos}}},
  \bibinfo{author}{\bibfnamefont{S.~L.} \bibnamefont{{Larson}}},
  \bibnamefont{et~al.}, \bibinfo{journal}{Classical and Quantum Gravity}
  \textbf{\bibinfo{volume}{34}}, \bibinfo{eid}{064003} (\bibinfo{year}{2017}),
  \eprint{1611.04596}.

\bibitem[{\citenamefont{George et~al.}(2017)\citenamefont{George, Shen, and
  Huerta}}]{George:2017fbn}
\bibinfo{author}{\bibfnamefont{D.}~\bibnamefont{George}},
  \bibinfo{author}{\bibfnamefont{H.}~\bibnamefont{Shen}}, \bibnamefont{and}
  \bibinfo{author}{\bibfnamefont{E.~A.} \bibnamefont{Huerta}}
  (\bibinfo{year}{2017}), \eprint{1706.07446}.

\bibitem[{\citenamefont{Wei and Huerta}(2019)}]{Wei:2019zlc}
\bibinfo{author}{\bibfnamefont{W.}~\bibnamefont{Wei}} \bibnamefont{and}
  \bibinfo{author}{\bibfnamefont{E.~A.} \bibnamefont{Huerta}}
  (\bibinfo{year}{2019}), \eprint{1901.00869}.

\bibitem[{\citenamefont{Shen et~al.}(2019{\natexlab{a}})\citenamefont{Shen,
  George, Huerta, and Zhao}}]{Shen:2019ohi}
\bibinfo{author}{\bibfnamefont{H.}~\bibnamefont{Shen}},
  \bibinfo{author}{\bibfnamefont{D.}~\bibnamefont{George}},
  \bibinfo{author}{\bibfnamefont{E.~A.} \bibnamefont{Huerta}},
  \bibnamefont{and} \bibinfo{author}{\bibfnamefont{Z.}~\bibnamefont{Zhao}}
  (\bibinfo{year}{2019}{\natexlab{a}}), \eprint{1903.03105}.

\bibitem[{\citenamefont{Zevin et~al.}(2017)\citenamefont{Zevin, Coughlin,
  Bahaadini, Besler, Rohani, Allen, Cabero, Crowston, Katsaggelos, Larson
  et~al.}}]{0264-9381-34-6-064003}
\bibinfo{author}{\bibfnamefont{M.}~\bibnamefont{Zevin}},
  \bibinfo{author}{\bibfnamefont{S.}~\bibnamefont{Coughlin}},
  \bibinfo{author}{\bibfnamefont{S.}~\bibnamefont{Bahaadini}},
  \bibinfo{author}{\bibfnamefont{E.}~\bibnamefont{Besler}},
  \bibinfo{author}{\bibfnamefont{N.}~\bibnamefont{Rohani}},
  \bibinfo{author}{\bibfnamefont{S.}~\bibnamefont{Allen}},
  \bibinfo{author}{\bibfnamefont{M.}~\bibnamefont{Cabero}},
  \bibinfo{author}{\bibfnamefont{K.}~\bibnamefont{Crowston}},
  \bibinfo{author}{\bibfnamefont{A.~K.} \bibnamefont{Katsaggelos}},
  \bibinfo{author}{\bibfnamefont{S.~L.} \bibnamefont{Larson}},
  \bibnamefont{et~al.}, \bibinfo{journal}{Classical and Quantum Gravity}
  \textbf{\bibinfo{volume}{34}}, \bibinfo{pages}{064003}
  (\bibinfo{year}{2017}),
  \urlprefix\url{http://stacks.iop.org/0264-9381/34/i=6/a=064003}.

\bibitem[{\citenamefont{Bahaadini et~al.}(2018)\citenamefont{Bahaadini,
  Noroozi, Rohani, Coughlin, Zevin, Smith, Kalogera, and
  Katsaggelos}}]{BAHAADINI2018172}
\bibinfo{author}{\bibfnamefont{S.}~\bibnamefont{Bahaadini}},
  \bibinfo{author}{\bibfnamefont{V.}~\bibnamefont{Noroozi}},
  \bibinfo{author}{\bibfnamefont{N.}~\bibnamefont{Rohani}},
  \bibinfo{author}{\bibfnamefont{S.}~\bibnamefont{Coughlin}},
  \bibinfo{author}{\bibfnamefont{M.}~\bibnamefont{Zevin}},
  \bibinfo{author}{\bibfnamefont{J.}~\bibnamefont{Smith}},
  \bibinfo{author}{\bibfnamefont{V.}~\bibnamefont{Kalogera}}, \bibnamefont{and}
  \bibinfo{author}{\bibfnamefont{A.}~\bibnamefont{Katsaggelos}},
  \bibinfo{journal}{Information Sciences} \textbf{\bibinfo{volume}{444}},
  \bibinfo{pages}{172 } (\bibinfo{year}{2018}), ISSN \bibinfo{issn}{0020-0255},
  \urlprefix\url{http://www.sciencedirect.com/science/article/pii/S0020025518301634}.

\bibitem[{\citenamefont{Allen et~al.}(2019)}]{Allen:2019dkq}
\bibinfo{author}{\bibfnamefont{G.}~\bibnamefont{Allen}} \bibnamefont{et~al.}
  (\bibinfo{year}{2019}), \eprint{1902.00522}.

\bibitem[{\citenamefont{Gabbard et~al.}(2019)\citenamefont{Gabbard, Messenger,
  Heng, Tonolini, and Murray-Smith}}]{Gabbard:2019rde}
\bibinfo{author}{\bibfnamefont{H.}~\bibnamefont{Gabbard}},
  \bibinfo{author}{\bibfnamefont{C.}~\bibnamefont{Messenger}},
  \bibinfo{author}{\bibfnamefont{I.~S.} \bibnamefont{Heng}},
  \bibinfo{author}{\bibfnamefont{F.}~\bibnamefont{Tonolini}}, \bibnamefont{and}
  \bibinfo{author}{\bibfnamefont{R.}~\bibnamefont{Murray-Smith}}
  (\bibinfo{year}{2019}), \eprint{1909.06296}.

\bibitem[{\citenamefont{Chatterjee et~al.}(2019)\citenamefont{Chatterjee, Wen,
  Vinsen, Kovalam, and Datta}}]{Chatterjee:2019gqr}
\bibinfo{author}{\bibfnamefont{C.}~\bibnamefont{Chatterjee}},
  \bibinfo{author}{\bibfnamefont{L.}~\bibnamefont{Wen}},
  \bibinfo{author}{\bibfnamefont{K.}~\bibnamefont{Vinsen}},
  \bibinfo{author}{\bibfnamefont{M.}~\bibnamefont{Kovalam}}, \bibnamefont{and}
  \bibinfo{author}{\bibfnamefont{A.}~\bibnamefont{Datta}}
  (\bibinfo{year}{2019}), \eprint{1909.06367}.

\bibitem[{\citenamefont{Miller et~al.}(2019)}]{Miller:2019jtp}
\bibinfo{author}{\bibfnamefont{A.~L.} \bibnamefont{Miller}}
  \bibnamefont{et~al.} (\bibinfo{year}{2019}), \eprint{1909.02262}.

\bibitem[{\citenamefont{Krastev}(2019)}]{Krastev:2019koe}
\bibinfo{author}{\bibfnamefont{P.~G.} \bibnamefont{Krastev}}
  (\bibinfo{year}{2019}), \eprint{1908.03151}.

\bibitem[{\citenamefont{Shen et~al.}(2019{\natexlab{b}})\citenamefont{Shen,
  Huerta, Zhao, Jennings, and Sharma}}]{Shen:2019vep}
\bibinfo{author}{\bibfnamefont{H.}~\bibnamefont{Shen}},
  \bibinfo{author}{\bibfnamefont{E.~A.} \bibnamefont{Huerta}},
  \bibinfo{author}{\bibfnamefont{Z.}~\bibnamefont{Zhao}},
  \bibinfo{author}{\bibfnamefont{E.}~\bibnamefont{Jennings}}, \bibnamefont{and}
  \bibinfo{author}{\bibfnamefont{H.}~\bibnamefont{Sharma}}
  (\bibinfo{year}{2019}{\natexlab{b}}), \eprint{1903.01998}.

\bibitem[{\citenamefont{Gabbard et~al.}(2018)\citenamefont{Gabbard, Williams,
  Hayes, and Messenger}}]{PhysRevLett.120.141103}
\bibinfo{author}{\bibfnamefont{H.}~\bibnamefont{Gabbard}},
  \bibinfo{author}{\bibfnamefont{M.}~\bibnamefont{Williams}},
  \bibinfo{author}{\bibfnamefont{F.}~\bibnamefont{Hayes}}, \bibnamefont{and}
  \bibinfo{author}{\bibfnamefont{C.}~\bibnamefont{Messenger}},
  \bibinfo{journal}{Phys. Rev. Lett.} \textbf{\bibinfo{volume}{120}},
  \bibinfo{pages}{141103} (\bibinfo{year}{2018}),
  \urlprefix\url{https://link.aps.org/doi/10.1103/PhysRevLett.120.141103}.

\bibitem[{\citenamefont{Wang et~al.}(2019)\citenamefont{Wang, Li, Yang, and
  Li}}]{Wang:2019ybv}
\bibinfo{author}{\bibfnamefont{L.-L.} \bibnamefont{Wang}},
  \bibinfo{author}{\bibfnamefont{J.}~\bibnamefont{Li}},
  \bibinfo{author}{\bibfnamefont{N.}~\bibnamefont{Yang}}, \bibnamefont{and}
  \bibinfo{author}{\bibfnamefont{X.}~\bibnamefont{Li}}, \bibinfo{journal}{New
  J. Phys.} \textbf{\bibinfo{volume}{21}}, \bibinfo{pages}{043005}
  (\bibinfo{year}{2019}).

\bibitem[{\citenamefont{Dreissigacker et~al.}(2019)\citenamefont{Dreissigacker,
  Sharma, Messenger, Zhao, and Prix}}]{PhysRevD.100.044009}
\bibinfo{author}{\bibfnamefont{C.}~\bibnamefont{Dreissigacker}},
  \bibinfo{author}{\bibfnamefont{R.}~\bibnamefont{Sharma}},
  \bibinfo{author}{\bibfnamefont{C.}~\bibnamefont{Messenger}},
  \bibinfo{author}{\bibfnamefont{R.}~\bibnamefont{Zhao}}, \bibnamefont{and}
  \bibinfo{author}{\bibfnamefont{R.}~\bibnamefont{Prix}},
  \bibinfo{journal}{Phys. Rev. D} \textbf{\bibinfo{volume}{100}},
  \bibinfo{pages}{044009} (\bibinfo{year}{2019}),
  \urlprefix\url{https://link.aps.org/doi/10.1103/PhysRevD.100.044009}.

\bibitem[{\citenamefont{{Morawski} et~al.}(2019)\citenamefont{{Morawski},
  {Bejger}, and {Ciecielag}}}]{2019arXiv190706917M}
\bibinfo{author}{\bibfnamefont{F.}~\bibnamefont{{Morawski}}},
  \bibinfo{author}{\bibfnamefont{M.}~\bibnamefont{{Bejger}}}, \bibnamefont{and}
  \bibinfo{author}{\bibfnamefont{P.}~\bibnamefont{{Ciecielag}}},
  \bibinfo{journal}{arXiv e-prints} \bibinfo{eid}{arXiv:1907.06917}
  (\bibinfo{year}{2019}), \eprint{1907.06917}.

\bibitem[{\citenamefont{Astone et~al.}(2018)\citenamefont{Astone,
  Cerd\'a-Dur\'an, Di~Palma, Drago, Muciaccia, Palomba, and
  Ricci}}]{PhysRevD.98.122002}
\bibinfo{author}{\bibfnamefont{P.}~\bibnamefont{Astone}},
  \bibinfo{author}{\bibfnamefont{P.}~\bibnamefont{Cerd\'a-Dur\'an}},
  \bibinfo{author}{\bibfnamefont{I.}~\bibnamefont{Di~Palma}},
  \bibinfo{author}{\bibfnamefont{M.}~\bibnamefont{Drago}},
  \bibinfo{author}{\bibfnamefont{F.}~\bibnamefont{Muciaccia}},
  \bibinfo{author}{\bibfnamefont{C.}~\bibnamefont{Palomba}}, \bibnamefont{and}
  \bibinfo{author}{\bibfnamefont{F.}~\bibnamefont{Ricci}},
  \bibinfo{journal}{Phys. Rev. D} \textbf{\bibinfo{volume}{98}},
  \bibinfo{pages}{122002} (\bibinfo{year}{2018}),
  \urlprefix\url{https://link.aps.org/doi/10.1103/PhysRevD.98.122002}.

\bibitem[{\citenamefont{Gebhard et~al.}(2019)\citenamefont{Gebhard, Kilbertus,
  Harry, and Scholkopf}}]{Gebhard:2019ldz}
\bibinfo{author}{\bibfnamefont{T.~D.} \bibnamefont{Gebhard}},
  \bibinfo{author}{\bibfnamefont{N.}~\bibnamefont{Kilbertus}},
  \bibinfo{author}{\bibfnamefont{I.}~\bibnamefont{Harry}}, \bibnamefont{and}
  \bibinfo{author}{\bibfnamefont{B.}~\bibnamefont{Scholkopf}}
  (\bibinfo{year}{2019}), \eprint{1904.08693}.

\bibitem[{\citenamefont{Vallisneri et~al.}(2015)\citenamefont{Vallisneri,
  Kanner, Williams, Weinstein, and Stephens}}]{Vallisneri_2015}
\bibinfo{author}{\bibfnamefont{M.}~\bibnamefont{Vallisneri}},
  \bibinfo{author}{\bibfnamefont{J.}~\bibnamefont{Kanner}},
  \bibinfo{author}{\bibfnamefont{R.}~\bibnamefont{Williams}},
  \bibinfo{author}{\bibfnamefont{A.}~\bibnamefont{Weinstein}},
  \bibnamefont{and} \bibinfo{author}{\bibfnamefont{B.}~\bibnamefont{Stephens}},
  \bibinfo{journal}{Journal of Physics: Conference Series}
  \textbf{\bibinfo{volume}{610}}, \bibinfo{pages}{012021}
  (\bibinfo{year}{2015}),
  \urlprefix\url{https://doi.org/10.1088%2F1742-6596%2F610%2F1%2F012021}.

\bibitem[{\citenamefont{Cao and Han}(2017)}]{PhysRevD.96.044028}
\bibinfo{author}{\bibfnamefont{Z.}~\bibnamefont{Cao}} \bibnamefont{and}
  \bibinfo{author}{\bibfnamefont{W.-B.} \bibnamefont{Han}},
  \bibinfo{journal}{Phys. Rev. D} \textbf{\bibinfo{volume}{96}},
  \bibinfo{pages}{044028} (\bibinfo{year}{2017}),
  \urlprefix\url{https://link.aps.org/doi/10.1103/PhysRevD.96.044028}.

\bibitem[{\citenamefont{Glorot and Bengio}(2010)}]{glorot2010understanding}
\bibinfo{author}{\bibfnamefont{X.}~\bibnamefont{Glorot}} \bibnamefont{and}
  \bibinfo{author}{\bibfnamefont{Y.}~\bibnamefont{Bengio}}, in
  \emph{\bibinfo{booktitle}{Proceedings of the thirteenth international
  conference on artificial intelligence and statistics}}
  (\bibinfo{year}{2010}), pp. \bibinfo{pages}{249--256}.

\bibitem[{\citenamefont{Kingma and Ba}(2014)}]{Kingma:2014vow}
\bibinfo{author}{\bibfnamefont{D.~P.} \bibnamefont{Kingma}} \bibnamefont{and}
  \bibinfo{author}{\bibfnamefont{J.}~\bibnamefont{Ba}} (\bibinfo{year}{2014}),
  \eprint{1412.6980}.

\bibitem[{\citenamefont{{Chen} et~al.}(2015)\citenamefont{{Chen}, {Li}, {Li},
  {Lin}, {Wang}, {Wang}, {Xiao}, {Xu}, {Zhang}, and
  {Zhang}}}]{2015arXiv151201274C}
\bibinfo{author}{\bibfnamefont{T.}~\bibnamefont{{Chen}}},
  \bibinfo{author}{\bibfnamefont{M.}~\bibnamefont{{Li}}},
  \bibinfo{author}{\bibfnamefont{Y.}~\bibnamefont{{Li}}},
  \bibinfo{author}{\bibfnamefont{M.}~\bibnamefont{{Lin}}},
  \bibinfo{author}{\bibfnamefont{N.}~\bibnamefont{{Wang}}},
  \bibinfo{author}{\bibfnamefont{M.}~\bibnamefont{{Wang}}},
  \bibinfo{author}{\bibfnamefont{T.}~\bibnamefont{{Xiao}}},
  \bibinfo{author}{\bibfnamefont{B.}~\bibnamefont{{Xu}}},
  \bibinfo{author}{\bibfnamefont{C.}~\bibnamefont{{Zhang}}}, \bibnamefont{and}
  \bibinfo{author}{\bibfnamefont{Z.}~\bibnamefont{{Zhang}}},
  \bibinfo{journal}{ArXiv e-prints}  (\bibinfo{year}{2015}),
  \eprint{1512.01274}.

\bibitem[{\citenamefont{Nitz et~al.}(2019)\citenamefont{Nitz, Capano, Nielsen,
  Reyes, White, Brown, and Krishnan}}]{Nitz_2019}
\bibinfo{author}{\bibfnamefont{A.~H.} \bibnamefont{Nitz}},
  \bibinfo{author}{\bibfnamefont{C.}~\bibnamefont{Capano}},
  \bibinfo{author}{\bibfnamefont{A.~B.} \bibnamefont{Nielsen}},
  \bibinfo{author}{\bibfnamefont{S.}~\bibnamefont{Reyes}},
  \bibinfo{author}{\bibfnamefont{R.}~\bibnamefont{White}},
  \bibinfo{author}{\bibfnamefont{D.~A.} \bibnamefont{Brown}}, \bibnamefont{and}
  \bibinfo{author}{\bibfnamefont{B.}~\bibnamefont{Krishnan}},
  \bibinfo{journal}{The Astrophysical Journal} \textbf{\bibinfo{volume}{872}},
  \bibinfo{pages}{195} (\bibinfo{year}{2019}),
  \urlprefix\url{https://doi.org/10.3847%2F1538-4357%2Fab0108}.

\bibitem[{GRB()}]{GRBGCN}
\bibinfo{howpublished}{The Gamma-ray Coordinates Network (GCN), TAN: Transient
  Astronomy Network. \url{https://gcn.gsfc.nasa.gov}}.

\end{thebibliography}
\end{document}